\documentclass{pasa}%

\usepackage[pdftex]{graphicx}
\usepackage{aas_macros}
\usepackage{multirow,comment,multicol}
\usepackage[titletoc,title]{appendix}
\usepackage[usenames,dvipsnames,svgnames,hyperref]{xcolor}
\usepackage{varwidth}
\usepackage{tikz}
\usetikzlibrary{positioning,calc,backgrounds}
\tikzstyle{line}=[draw]
\tikzstyle{arrow}=[draw, -latex] 
\usepackage{tcolorbox}
\usepackage[
    pdftex,% get better results and can break links but can't use psfrag
    pdfpagelabels,
    breaklinks=true,
    bookmarks=true,
    bookmarksopen=false,
    bookmarksopenlevel=2
    bookmarksnumbered=true,
    bookmarkstype=toc,
    colorlinks=true,
    citecolor=RoyalBlue,
    linkcolor=ForestGreen,
    menucolor=Teal,
    urlcolor=DarkOrange,
    draft=true
]{hyperref}
\usepackage{subcaption}
\usepackage{amsfonts}
\usepackage{longtable}

\newcommand{\Eqref}[1]{Eq.~(\ref{#1})}
\newcommand{\Figref}[1]{Fig.~\ref{#1}}
\newcommand{\Secref}[1]{\S\ref{#1}}  
\newcommand{\Tableref}[1]{Table~\ref{#1}}

\def \Mpch{h^{-1}{\rm Mpc}}
\def \kpch{h^{-1}{\rm kpc}}
\def \Msunh{h^{-1}{\rm M}_\odot}

\title[\textsc{TreeFrog}]{Climbing Halo Merger Trees with \textsc{TreeFrog}}

\author[Elahi et al.]{
Pascal J. Elahi$^{1,2,\dagger}$, 
Rhys J.J. Poulton$^{1,2}$, 
Rodrigo J. Tobar$^{1}$, 
Rodrigo Ca\~nas$^{1,2}$, 
Claudia del P. Lagos$^{1,2}$, 
Chris Power$^{1,2}$,
Aaron S.~G. Robotham$^{1}$,
\affil{
$^1$International Centre for Radio Astronomy Research, University of Western Australia, 35 Stirling Highway, Crawley, WA 6009, Australia}
\affil{
$^2$ARC Centre of Excellence for All Sky Astrophysics in 3 Dimensions (ASTRO 3D)
}
\affil{$^\dagger$Email: pascal.elahi@icrar.org}
}%

\jid{PASA}
\doi{10.1017/pas.\the\year.xxx}
\jyear{\the\year}

\begin{document}

\begin{frontmatter}
\maketitle

\begin{abstract}
We present \textsc{TreeFrog}, a massively parallel \textit{halo merger tree builder} that is capable comparing different halo catalogues and producing halo merger trees. The code is written in \textsc{c++11}, use the MPI and OpenMP API's for parallelisation, and includes python tools to read/manipulate the data products produced. The code correlates binding energy sorted particle ID lists between halo catalogues, determining optimal descendant/progenitor matches using multiple snapshots, a merit function that maximises the number of shared particles using pseudo-radial moments, and a scheme for correcting halo merger tree pathologies. Focusing on \textsc{VELOCIraptor} catalogues for this work, we demonstrate how searching multiple snapshots spanning a dynamical time significantly reduces the number of stranded halos, those lacking a descendant or a progenitor, critically correcting poorly resolved halos. We present a new merit function that improves the distinction between primary and secondary progenitors, reducing tree pathologies. We find FOF accretion rates and merger rates show similar mass ratio dependence. The model merger rates from \cite{poole2017a} agree with the measured net growth of halos through mergers. 
%Total merger counts are converged at the level of ̃5 per cent for friends-of-friends (FoF) haloes of size np ≳ 75 across a factor of 512 in mass resolution, but substructure rates converge more slowly with mass resolution, reaching convergence of ̃10 per cent for np ≳ 100 and particle mass mp ≲ 109 M☉. We present analytic fits to FoF and substructure merger rates across nearly all observed galactic history (z ≤ 8.5). While we find good agreement with the results presented by Fakhouri et al. for FoF haloes, a slightly flatter dependence on merger ratio and increased major merger rates are found, reducing previously reported discrepancies with extended Press-Schechter estimates. When appropriately defined, substructure merger rates show a similar mass ratio dependence as FoF rates, but with stronger mass and redshift dependencies for their normalization. 
\end{abstract}

\end{frontmatter}

\section{Introduction}
\label{sec:intro}
Cosmological simulations underpin theoretical predictions of the formation and evolution of both galaxies and dark matter halos. Simulations containing billions of tracers are now common place, both N-body \cite[e.g., Millennium, MultiDark, TIAMAT, SURFS][]{springel2005,boylankolchin2009,klypin2016a,poole2016a,elahi2018a} and full hydrodynamical simulations \cite[e.g., EAGLE, ILLUSTRIS, Horizon-AGN][]{schaye2015a,vogelsberger2014a,dubois2014a}. These simulations are processed by sophisticated (sub)halo finders to identify dark matter (sub)halos \cite[see][for a discussion of (sub)halo finding]{knebe2011,onions2012,knebe2013a} and synthetic galaxies \cite[e.g.][]{canas2018a}. The evolution across cosmic time of galaxies and cosmic structure are reconstructed through the use of so-called ``tree builders''. Following the mass accretion history of dark matter halos and producing ``halo merger trees'', for instance, is pivotal in producing synthetic galaxy surveys with Semi-Analytic Models (SAM) of galaxy formation (e.g., \citealp{cole2000,knebe2018a,lagos2018a,baugh2018a}, though SAMs can also use extended Press-Schetcher theory to produce Monte Carlo trees calibrated against simulations, e.g., \citealp{parkinson2008a,benson2017a}). The role of ``Tree Builders'' is to identify the optimal descendants and progenitors of halos/galaxies found at a given snapshot to later and previous snapshots respectively. 

\par 
There are a variety of tree builders in use \cite[e.g.][]{behroozi2013b,dhalos,poole2017a}, most of which perform similarly well, at least for reconstructing the accretion history of field halos \cite[see][for an overview of tree building]{srisawat2013}. The problem of identifying optimal descendants/progenitors is compounded by imperfect (sub)halo finding as all (sub)halo finders can momentarily lose subhalos \cite[][]{avila2014a}. The cadence of the input halo catalogues can also have a severe impact on the resulting merger history, particularly when coupled with imperfect (sub)halo finding as (sub)halos can flicker in and out of existence \cite[][]{wang2016a}. The performance of merger trees can impact the synthetic galaxy population produced by SAMs \cite[][]{lee2014a}, though, in practice, only the satellite galaxy population is severely effect by flaws in halo merger trees. \cite{srisawat2013} identified three features Tree Builders should employ in some fashion: the use of particle IDs to match objects between snapshots; using multiple snapshots to identify matches; and a method to smooth out any large mass fluctuations.

\par 
Here we present \textsc{TreeFrog}, a halo merger tree builder that employs the first two most critical features outlined in \cite{srisawat2013}\footnote{Freely available \href{https://github.com/pelahi/TreeFrog.git}{\url{https://github.com/pelahi/TreeFrog.git}}. Documentation is found at \href{https://treefrog-halo-merger-tree-builder.readthedocs.io/en/latest/}{\url{https://treefrog-halo-merger-tree-builder.readthedocs.io/en/latest/}}}. When combined with state-of-the-art (sub)halo finders like \textsc{VELOCIraptor} \cite[][]{velociraptorpaper}, also minimises mass fluctuations of orbiting subhalos. The original, highly simplified \textsc{TreeFrog} algorithm was first briefly presented in \cite{srisawat2013}. Here we present significant updates and a full description of the code. 
 
\par
Our paper is organised as follows: in section \Secref{sec:treefrog}, we outline the code package, present tests of our algorithm in \Secref{sec:results:treefrog} and conclude in \Secref{sec:discussion} with a summary and discussion.
%For examples of there use, see \Secref{sec:appendix:examples}.

\section{Following the evolution of structure with TreeFrog}
\label{sec:treefrog}
\textsc{TreeFrog} at the most basic level is a particle correlator, matching particles present in one catalogue with those in another using particle IDs. It relies on particle IDs being continuous across time (or halo catalogues), so any particle type which has fluid IDs cannot be used to build a tree or cross-catalogue\footnote{An example would be the use of gas cells from AMR codes.}. The basics of the particle correlator was first introduced in \cite{srisawat2013}. Here we present the software in full and significant updates to the original code used in \cite{srisawat2013}. Readers interested in tests and results can skip to \Secref{sec:results:treefrog}. 

\par
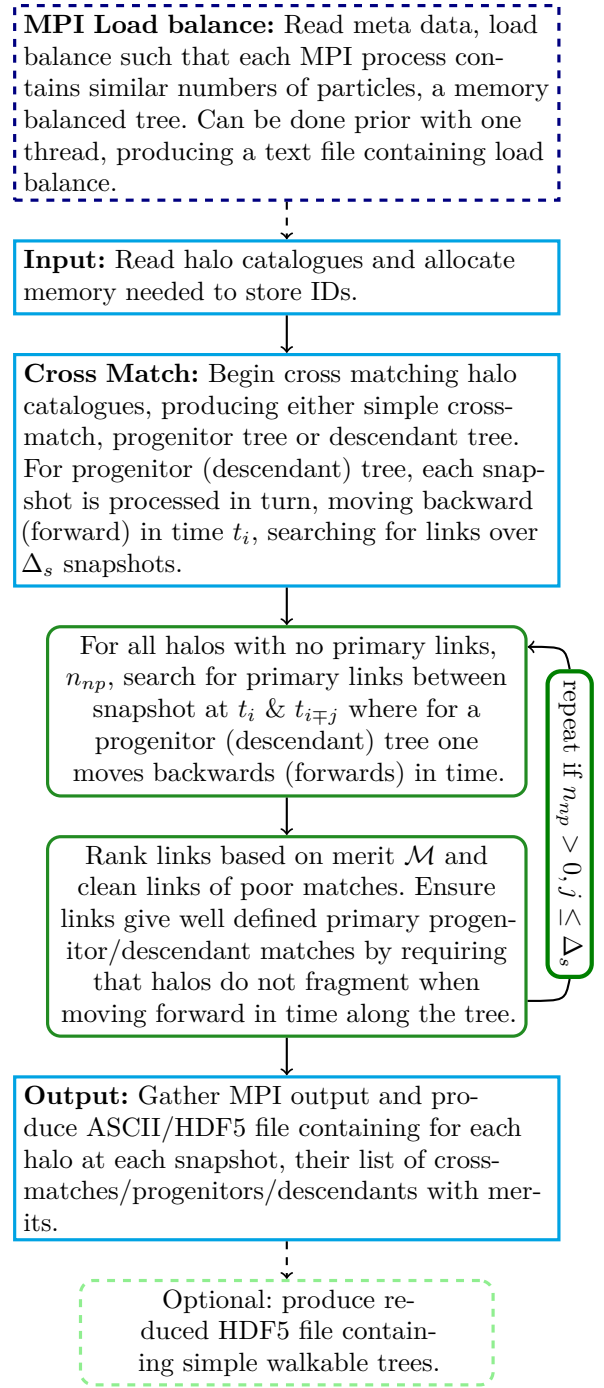
\begin{figure}
\centering
\begin{tikzpicture}[
    roundnode/.style={rectangle, rounded corners = 5pt, draw=ForestGreen, very thick, text width=0.35\textwidth,align=center},
    roundnode2/.style={rectangle, rounded corners = 5pt, draw=LightGreen, very thick, text width=0.3\textwidth,align=center},
    squarednode/.style={rectangle, draw=Cerulean, very thick, text width=0.4\textwidth, align=left},
    node distance=5mm
    ]
    %Nodes
    \node[squarednode, draw=NavyBlue, dashed](mpi){
    {\bf MPI Load balance:}
    Read meta data, load balance such that each MPI process contains similar numbers of particles, a memory balanced tree. Can be done prior with one thread, producing a text file containing load balance. 
    };
    \node[squarednode](input)[below=of mpi]{
    {\bf Input:}
    Read halo catalogues and allocate memory needed to store IDs. 
    };
    \node[squarednode](match)[below=of input]{
    {\bf Cross Match:}
    Begin cross matching halo catalogues, producing either simple cross-match, progenitor tree or descendant tree. For progenitor (descendant) tree, each snapshot is processed in turn, moving backward (forward) in time $t_i$, searching for links over $\Delta_s$ snapshots. 
    };
    \node[roundnode](loop1start)[below=of match]{
    For all halos with no primary links, $n_{np}$, search for primary links between snapshot at $t_i$ \& $t_{i\mp j}$  where for a progenitor (descendant) tree one moves backwards (forwards) in time.
    };
    \node[roundnode](loop1end)[below=of loop1start]{
    Rank links based on merit $\mathcal{M}$ and clean links of poor matches. Ensure links give well defined primary progenitor/descendant matches by requiring that halos do not fragment when moving forward in time along the tree. 
    };
    \node[rectangle,  rounded corners = 5pt, draw=Green, ultra thick, rotate=-90, yshift=15pt](loop1criterion)[below right=of loop1start, anchor=north]{
    \begin{varwidth}{0.22\textwidth}
    repeat if $n_{np}>0,j\leq\Delta_s$
    \end{varwidth}
    };
    \node[squarednode](output)[below=of loop1end]{
    {\bf Output:}
    Gather MPI output and produce ASCII/HDF5 file containing for each halo at each snapshot, their list of cross-matches/progenitors/descendants with merits. 
    };
    \node[roundnode2, dashed](additionaloutput)[below=of output]{
    Optional: produce reduced HDF5 file containing simple walkable trees.
    };
    
    %lines
    \draw[thick,dashed,->] (mpi.south) -- (input.north);
    \draw[->,thick] (input.south) -- (match.north);
    \draw[->,thick] (match.south) -- (loop1start.north);
    \draw[->,thick] (loop1start.south) -- (loop1end.north);
    \draw[->,thick] (loop1end.south) -- (output.north);
    \draw[->,thick,dashed] (output.south) -- (additionaloutput.north);

    %outer loop
    \draw[-,thick] (loop1end.345) .. controls +(right:5mm) .. (loop1criterion.east);
    \draw[<-,thick] (loop1start.15) .. controls +(right:5mm) .. (loop1criterion.west);

\end{tikzpicture}
    \caption{Activity chart for \textsc{TreeFrog}.}
    \label{fig:treefrog:scheme}
\end{figure}

 
The code can produce a simple cross catalogue or a full halo merger tree. When building a simple cross-catalogue, one reference catalogue is compared to another and all matches with high significance are returned. Full halo merger trees that try to capture the evolution of cosmic structure across cosmic time require extra care and can be constructed in two 
different fashion, either by walking backwards in time, a so-called progenitor based tree, or walking forwards in time, a so-called descendant based tree. A flow-chart of the code is presented in \Figref{fig:treefrog:scheme} and we describe the various aspects of our code below. For readers interested in input interfaces, output, and general modes of operation we suggest skipping to \Secref{sec:treefrog:summary}.
%and by default do not allow for halo fragmentation, i.e., objects can have multiple progenitors but only a single primary descendant

\subsection{Merit Function \& Optimal Matches} 
\label{sec:treefrog:merit}
The first step in producing a cross comparison of halo catalogues or full halo merger trees is the cross matching of particles in (sub)halos. The cross-match between catalogues $A$ \& $B$ is produced by identifying for each object in catalogue $A$, the object in catalogue $B$ that maximises a merit function. Several merit functions are available. For a simple comparison between two halo catalogues, the merit is defined as:
\begin{equation}
    \mathcal{N}_{A_{i}B_{j}} = N_{A_{i}\bigcap B_{j}}^2/(N_{A_{i}}N_{B_{j}}),\label{eqn:nshared}
\end{equation}
where $N_{A_{i}\bigcap B_{j}}$ is the number of particles shared between objects $i$ in catalogue $A$ and $j$ in catalogue $B$, and $N_{A_{i}}$ \& $N_{B_{j}}$ are the total number of particles in the respective objects. This merit function maximises the fraction of shared particles in both objects and is quite robust \cite[][]{libeskind2011b}. 

\par 
However, there are instances where several possible candidates are identified. This is particularly problematic in multi-merger events that naturally arise when constructing halo merger trees. During similar mass mergers, loosely bound particles can be readily exchanged between halos. 

\par 
We follow \cite{poole2017a} to alleviate these issues by using the rank of particles as ordered by their binding energy:
\begin{equation}
    \mathcal{S}_{A_{i}B_{j},A_{i}} = \sum_{l}^{N_{A_{i}\bigcap B_{j}}} 1/\mathcal{R}_{l,A_{i}}\label{eqn:ranking}
\end{equation}
Here the sum is over all shared particles and $\mathcal{R}_{l,A_{i}}$ is the rank of particle $l$ in halo $A_{i}$, with the most bound particle in the halo having $\mathcal{R}=1$. The maximum value, when all particles are shared, is $\mathcal{S}^{\rm max}_{A_{i}B_{j},A_{i}}=\gamma+\ln N_{A_{i}}$, with $\gamma=0.5772156649$ being the Euler-Mascheroni constant. 

\par 
This second merit requires input catalogues to be ordered according to binding energy\footnote{Technically, input catalogues need to be  sorted in a physically meaningful fashion for the desired comparison. For halos, radial sorting is also reasonable.} as \textsc{TreeFrog} does not calculate a ranking. \textsc{VELOCIraptor} natively has this ranking in place. Catalogues produced by other halo finders must be similarly sorted, otherwise it is strongly advised that one does not use this additional ranking merit in \Eqref{eqn:ranking}.

\par
We combine \Eqref{eqn:nshared} with the normalised version of \Eqref{eqn:ranking}, i.e $\tilde{\mathcal{S}}_{A_{i}B_{j},A_{i}}=\mathcal{S}_{A_{i}B_{j},A_{i}}/\mathcal{S}^{\rm max}_{A_{i}B_{j},A_{i}}$, to obtain 
\begin{equation}
    \mathcal{M}_{A_{i}B_{j}}=\mathcal{N}_{A_{i}B_{j}}\tilde{\mathcal{S}}_{A_{i}B_{j},A_{i}}\tilde{\mathcal{S}}_{A_{i}B_{j},B_{i}}, \label{eqn:merit}
\end{equation}
where we calculate the rank ordering for both halos in question since this ordering can be quite different. This combined merit maximises the total shared number of particles while also weighting the match by the number of equally well bound shared particles. 

\par 
Finally, not all particles need be used to calculate merits, particularly if the input catalogues are sorted in a physically meaningful way, such as binding energy. One can limit the merit to a fraction $f_{\rm TF}$ of these particles. Limiting the comparison to the most bound particles can be key to correctly following major mergers. 

\subsection{Halo Merger Trees}
\label{sec:treefrog:tree}
Halo merger trees are more than a simple cross comparison of halo catalogues. They follow the evolution of halos, the mass accretion history, tidal disruption and interaction with other halos, ideally following the formation of a halo across cosmic time till either the present day or the point at which the object is tidally disrupted as it falls into a larger halo. Before discussing how trees are constructed, it is important to lay out some terminology. 
\begin{itemize}
    \item A progenitor/descendant is a (sub)halo present at a previous/later time that points to a halo present at a later/earlier time as being its descendant/progenitor. 
    \item A primary progenitor/descendant is where a (sub)halo's descendant/progenitor points back to it as its progenitor/descendant. Note that due to issues with the halo finding process and physical processes involving mass transfer, it is possible for objects to have more than a single candidate descendant. We discuss this in more detail later. 
    \item A main branch in a tree is one which traces the evolution of a halo from its first progenitor to its final descendant, moving forwards/backwards via the object's primary descendant/primary progenitor links. 
    \item A secondary progenitor is where a (sub)halo has merged with the main branch of another halo and ceases to exist as an independent object. This defines sub-branches of the main branch and is a natural consequence of structure formation which we discuss in more detail later. 
    \item A secondary descendant is where a (sub)halo has identified multiple possible links. All links that have lower merits are secondary matches. Such matches are a natural consequence of mass loss, where an object falls into another object and still has a surviving primary descendant but some of its mass has been associated with the accreting object, generating a low merit link. 
    \item The first (root) progenitor of a branch is the object that has no progenitors. The final (root) descendant of a branch is the object that has no descendants, which typically occurs at the last snapshot. 
\end{itemize}

\par 
Building trees is complicated by the imperfect (sub)halo finding process \cite[see][for discussions of the pitfalls of tree building]{srisawat2013,behroozi2013b,avila2014a,wang2016a,poole2017a}. Halo finders can artificially merge halos at a given snapshot and later separate them, generating missing links in the tree. This problem is particularly acute for low mass halos that lie near the particle number threshold used by the halo finder or for subhalos close to the centre of their host halo. Stranded (sub)halos lacking a progenitor are less of a problem with \textsc{VELOCIraptor}, the (sub)halo finder used in this study, but no (sub)halo finder is completely immune. Critically, such events can occur at much higher masses. With fine-scale temporal resolution, the merger tree will have halos popping in and out of existence. 

\par
This can lead to several crucial issues in the resulting tree: 
\begin{itemize}
    \item Truncation: where the (sub)halo finder cannot find an object for one or more snapshots, leading to premature disruption of the object, and possibly leaving a large object identified at later times with no progenitor (leading to the appearance of halo fragmentation).
    \item Flip-flopping: where links between two objects are swapped at one snapshot but corrected in subsequent snapshot(s), leading to large changes in the object's properties in the snapshot where it happens.
    \item Branch swapping: which is similar to flip-flopping, except the tree builder does not correct it and so the objects continue their independent evolution, leaving a single point with a sharp change in properties. 
\end{itemize}
All trees will suffer from these issues, the degree to which they do depending in part on the (sub)halo finder. A variety of methods have been used in literature to handle this cases, some simple \citep[e.g.][which are specific for simple FOF merger trees]{fakhouri2008a,genel2010a}, some more complex \cite[e.g.][]{behroozi2013b,poole2017a}. \textsc{TreeFrog} uses several techniques to minimise the occurrence of these issues, the most critical one being that it searches multiple snapshots for candidate links. We discuss the specific extra steps taken when producing progenitor and descendant based tree. 

\paragraph{Progenitor Based Tree:}
The input catalogue is processed by comparing objects (both halos and subhalos) found at a snapshot to those in preceding snapshots, moving backwards in time. We start by linking a snapshot with the one immediately preceding it, identifying matches for all objects. Objects are allowed to have multiple progenitors but no object by construction will have multiple descendants. We rank temporal links such that the primary progenitor of an object is the one which maximises the merit looking backwards in time. If two objects share the same progenitor, the one with the lower merit has the link removed. If an object has a poor merit, typically below $\mathcal{M}_{\rm lim}\sim0.05$, the link is removed. The remaining, highest merit link is deemed a primary progenitor/descendant link. For objects with no progenitor, earlier snapshots are searched until a viable progenitor is found, up to a maximum number of snapshots $\Delta_s$ from the current one. 

\paragraph{Descendant Based Tree:}
The input catalogue is processed by comparing objects (both halos and subhalos) found at a snapshot to those found at later snapshots, moving forwards in time. We start by linking a snapshot with the one immediately following it, identifying matches for all objects. In the absence of tidal disruption, objects should have a single descendant. However, mass loss and tidal disruption are natural processes that complicated tree building, producing links to several candidate descendants. Consequently, we allow objects to have multiple candidate descendants. Candidate descendants are split into two categories: primary and secondary links. A primary descendant link is one where a halo's best candidate descendant, that is the one with highest merit amongst the object's matches, also ranks the halo as the best amongst all its matches going backward in time. All other connections are classified as secondary links. Secondary links arise from the physical tidal disruption and merging processes as well as unphysical merging of halos where the halo finder fails to identify an object. \textsc{TreeFrog} does not attempt to differentiate between these processes. If an object does not have a primary descendant, subsequent snapshots are searched till a primary is identified or the maximum number of snapshots to be searched, $\Delta_s$, has been reached. 

\par
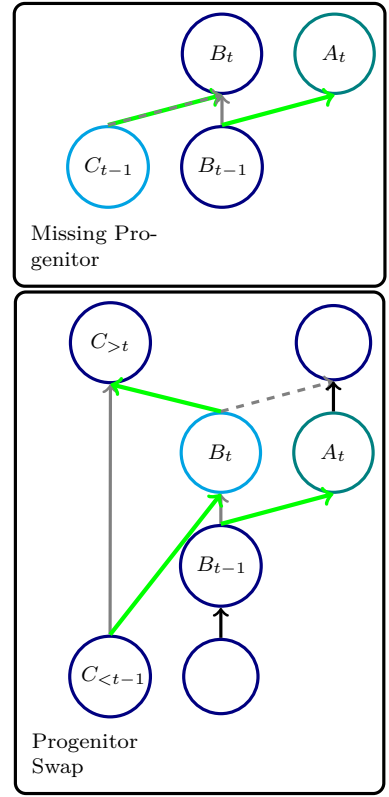
\begin{figure}
\centering
\begin{tikzpicture}[framed, background rectangle/.style={very thick,draw=black, rounded corners},
    labelnode/.style={rectangle, text width=50pt,align=left},
    halonode/.style={circle, draw=Cerulean, very thick, text width=22pt, align=center},
    node distance=4mm,
    ]
    \footnotesize
    %Nodes
    \node[halonode, draw=NavyBlue](haloB1){
    $B_{t-1}$
    };
    \node[halonode, draw=Cerulean,left=of haloB1](haloC1){
    $C_{t-1}$
    };
    \node[halonode, draw=NavyBlue,above=of haloB1](haloB2){
    $B_{t}$
    };
    \node[halonode, draw=Teal](haloA1)[right=of haloB2]{
    $A_{t}$
    };
   
    %lines
    \draw[->,very thick,gray] (haloB1.north) -- (haloB2.south);
    \draw[->,ultra thick,green] (haloC1.north) -- (haloB2.south);
    \draw[->,ultra thick,green] (haloB1.north) -- (haloA1.south);
    \draw[->,very thick,dashed,gray] (haloC1.north) -- (haloB2.south);
    
    \node[labelnode](label)[below left=of haloB1]{Missing Progenitor};
\end{tikzpicture}
\begin{tikzpicture}[framed, background rectangle/.style={very thick,draw=black, rounded corners},
    labelnode/.style={rectangle, text width=50pt,align=left},
    halonode/.style={circle, draw=Cerulean, very thick, text width=22pt, align=center},
    node distance=4mm
    ]
    \footnotesize
    %Nodes
    \node[halonode, draw=NavyBlue](haloB1){
    $B_{t-1}$
    };
    \node[halonode, draw=Cerulean,above=of haloB1](haloB2){
    $B_{t}$
    };
    \node[halonode, draw=White](haloB3)[above=of haloB2]{
    };
    \node[halonode, draw=NavyBlue](haloB0)[below=of haloB1]{
    };

    \node[halonode, draw=NavyBlue,left=of haloB0](haloC1){
    $C_{<t-1}$
    };
    \node[halonode, draw=NavyBlue,left=of haloB3](haloC2){
    $C_{>t}$
    };
   
    \node[halonode, draw=Teal](haloA1)[right=of haloB2]{
    $A_{t}$
    };
    \node[halonode, draw=NavyBlue](haloA2)[above=of haloA1]{
    };

    %lines
    \draw[->,very thick,gray] (haloB1.north) -- (haloB2.south);
    \draw[->,very thick,gray] (haloC1.north) -- (haloC2.south);
    \draw[->,very thick,dashed,gray] (haloB2.north) -- (haloA2.south);
    \draw[->,ultra thick,green] (haloC1.north) -- (haloB2.south);
    \draw[->,ultra thick,green] (haloB1.north) -- (haloA1.south);
    \draw[->,ultra thick,green] (haloB2.north) -- (haloC2.south);
    \draw[->,very thick,black] (haloB0.north) -- (haloB1.south);
    \draw[->,very thick,black] (haloA1.north) -- (haloA2.south);

    \node[labelnode](label)[below left=of haloB0]{Progenitor Swap};
\end{tikzpicture}
    \caption{{\bf Branch Fixes:} Diagrams show branch fixes. Original primary and secondary progenitors are highlighted by navy blue and light blue circles. Halo missing a progenitor is in teal. Solid and dashed lines connect primary and secondary progenitors respectively. Original connections are in gray, new connections in green.}
    \label{fig:treefrog:branchfixes}
\end{figure}

 
Once an initial tree has been constructed, \textsc{TreeFrog} attempts to correct the tree for truncation events (and the associated branch swapping that may result from them) that arise from the loss of the object by the (sub)halo finder. Objects that lack a primary progenitor are corrected for as follows. For an object $A_t$ that does not have a primary progenitor, we examine the best ranked secondary progenitor, $B_{t-1}$ and this secondary progenitor's primary descendant, $B_t$, if such an object exists. If this object $B_t$ has a secondary progenitor $C_{t-1}$ which itself has no primary descendant and has a merit $\mathcal{M}_{C_{t-1}B_{t}}$ within a factor of $f_{\mathcal{M}}\sim0.5$ of $\mathcal{M}_{B_{t-1}B_{t}}$ and above $\mathcal{M}_{\rm lim}$, we adjust the links so that instead of $(B_{t-1}\rightarrow B_t, C_{t-1}\rightarrow\varnothing,\varnothing\rightarrow A_{t})$, we have $(C_{t-1}\rightarrow B_t,B_{t-1}\rightarrow A_t)$. A schematic of this branch fix is shown in the top panel of \Figref{fig:treefrog:branchfixes}. 

\par
We also apply further corrections to objects with no primary progenitor as a post-processing step that relies on using the full history of an object, specifically the final descendant of a main branch. We identify objects that do not have primary progenitors but have secondary progenitors. Specifically, for an object $A_t$, we examine its best ranked secondary progenitor, $B_{t-1}$, and that object's best descendant ${B_t}$. If both objects end up with the same final descendant, it is possible progenitors have been incorrectly assigned due to artificial merging of objects at $t-1$. Thus we then search for an object $C_{t_i<t-1}$ that has a primary descendant after the merger at $t-1$, $C_{t_i>t}$, that belongs to the same final descendant and that has a similar phase-space position and number of particles as $B_{t}$. Specifically, we require:
\begin{align}
    ({\bf x}_{C_{t<t-1}}-{\bf x}_{B_{t}})/R(V_{\rm max})_{B_{t}},
        &\leq\alpha_{R(V_{\rm max})} \notag\\
    ({\bf v}_{C_{t<t-1}}-{\bf v}_{B_{t}})/V_{{\rm max},B_{t}},
        &\leq\alpha_{V_{\rm max}}  \notag\\
    N_{C_{t<t-1}}&\geq \alpha_{N} N_{B_{t}}.
\end{align}
Here ${\bf x}$ \& ${\bf v}$ are the positions \& velocities, $R(V_{\rm max})$ \& $V_{\rm max}$, are the maximum circular velocity radius \& is the maximum circular velocity and $N$ is the number of particles belonging to the object. This phase-space check is a simplified form of halo tracking. We also limit this patching to well resolved objects, that is those composed of $\beta_{N_{\rm lim}}N_{\rm lim}$ where $N_{\rm lim}$ is the particle limit used by the (sub)halo finder, $\beta_{N_{\rm lim}}\gtrsim2$. Full gravitational evolution and unbinding, particularly for poorly resolved objects, is best done using halo tracking tools (like WhereWolf, Poulton et al., in prep). If this match meets these criteria, we then correct the branches so that instead of $(B_{t-1}\rightarrow B_t, C_{<t-1}\rightarrow C_{>t},\varnothing\rightarrow A_t)$, we have $(B_{t-1}\rightarrow A_t, C_{<t-1}\rightarrow B_{t}\rightarrow C_{>t})$. A schematic of this branch fix is shown in the bottom panel of \Figref{fig:treefrog:branchfixes}. The parameters $\alpha$ are order unity and we have found $(\alpha_{R(V_{\rm max})}, \alpha_{V_{\rm max}}, \alpha_N)=(2.0,1.0,0.05)$ corrects most events.

\subsection{Code Structure}
\label{sec:treefrog:summary}
\textsc{TreeFrog} is a {\sc c++} code that uses OpenMP+MPI APIs for parallelisation but can be compiled in serial mode, solely with OpenMP, and solely with MPI. The code requires an input file containing a list of halo catalogue file names, the number of snapshots to process and an output file name.

\par 
The main input file is a simple text file that lists the locations of the halo catalogues ordered in increasing time.
\begin{comment} 
i.e., 
\begin{packeditemize}
    \item[] snapshot\_000
    \item[] snapshot\_001
    \item[] snapshot\_002
    \item[] $\vdots$
    \item[] snapshot\_nnn
\end{packeditemize}
\end{comment}
These halo catalogues can be in native \textsc{VELOCIraptor} output (ASCII, Binary, and HDF5\footnote{Self-describing binary format, library found at \href{https://www.hdfgroup.org/}{https://www.hdfgroup.org/}}) or in a simple ASCII format that was used in the SUSSING Merger Trees workshop \cite[see][]{srisawat2013}. Other input formats can be implemented and the input data must contain a list of particle IDs and possibly particle types for each halo in the halo catalogue.

\par 
MPI domain decomposition is temporal in nature, with each thread loading the halos from an entire simulation volume for a certain number of snapshots. Snapshots are distributed to different threads by load balancing the memory footprint on each MPI process. Specifically, snapshots are split to ensure that each thread loads either roughly the same number of total halos or the same total number of particle ids (depending on runtime configuration). An MPI thread loads a minimum of $2\Delta_s+1$ and each MPI domain must overlap the neighbouring domains by $\Delta_s$ so as to have a complete list of connections to and from the snapshots localised to a single mpi domain. 

\par 
Although there are a variety of modes that \textsc{TreeFrog} can be operated in, there are three principal ones. \textsc{TreeFrog} can be used to produce a Descendant Tree, Progenitor Tree or simply cross correlate two catalogues. It produces the following types of output formats: ASCII; {\sc HDF5} ({\em preferred}). The output file(s) consists of a list of a halo, the number of descendant/progenitor/cross matches and the ID of these linked objects, for all halos identified at a given snapshot. In the ASCII format, this is combined into a single continuous file, whereas in the preferred HDF5 format, each snapshot is written separately. An ADIOS interface will be included and will have the same naming convention as the HDF5 output. 

\par 
Post-processing of the full tree information is done using python scripts to produce a simple, walkable tree where each (sub)halo will have links to their immediate progenitor, immediate descendant, first progenitor and final descendant (that is eliminating all secondary links and merit information), the typical information need by SAMs. This post-processing removes secondary links. 

\par 
Options can be passed either via command line or through a text file. We list the configuration options that can be passed via this input text file in \Tableref{tab:treefrog:params}.
\begin{table*}
\setlength\tabcolsep{2pt}
\centering\footnotesize
\caption{Key \textsc{TreeFrog} parameters}
\label{tab:treefrog:params}
\begin{tabular}{@{\extracolsep{\fill}}p{0.125\textwidth}|p{0.23\textwidth}|c|p{0.49\textwidth}}
\hline
\hline
    & Name & Default Value & Comments\\
    \hline
    General Tree Options & & & Related to general tree construction. \\
    \hline
    & Tree\_direction & 1
        & Integer indicating direction in which to process snapshots and build the tree. Descendant [1], Progenitor [0], or Both [-1].\\
    & Particle\_type\_to\_use & -1
        & Particle types to use when calculating merits. All [-1], Gas [0], Dark Matter [1], Star [4].\\
    & Default\_values& 1
        & Whether to use default cross matching \& merit options when building the tree. 1/0 for True/False.\\
    
    \hline
    Merit Options & & & Related to calculation of merit function. \\
    \hline
    & Merit\_type & 6  
        & Integer specifying merit function to use. Optimal descendant tree merit in \Eqref{eqn:merit} [6], common (progenitor tree) merit in \Eqref{eqn:nshared} [1].\\
    & Core\_match\_type & 2
        & Integer flag indicating the type of core matching used. Off [0], core-to-all [1], core-to-all followed by core-to-core [2], core-to-core only [3].\\
    & Particle\_core\_fraction & 0.4
        & Fraction of particles to use when calculating merits. Assumes some meaningful rank ordering to input particle lists and uses the first $f_{\rm TF}$ fraction.\\
    & Particle\_core\_min\_numpart & 5
        & Minimum number of particles to use when calculating merit if core fraction matching enabled.\\

    \hline
    Temporal Linking Options & & & Related to how code searches for candidate links across multiple snapshots.\\
    \hline
    & Nsteps\_search\_new\_links & 1
        & Number of snapshots to search for links.\\
    & Multistep\_linking\_criterion & 3
        & Integer specifying the criteria used when deciding whether more snapshots should be searched for candidate links. Criteria depend on tree direction. {\bf Descendant Tree:} continue searching if halo is: missing descendant [0]; missing descendant or descendant merit is low [1]; missing descendant or missing primary descendant [2]; missing a descendant, a primary descendant or primary descendant has poor merit [3]. {\bf Progenitor tree}: [0,1]. \\
    & Merit\_limit\_continuing\_search & 0.025 
        & Float specifying the merit limit a match must meed if using Multistep\_linking\_criterion=[1,3].\\

\end{tabular}
\end{table*}

\begin{comment}
\begin{table*}
\setlength\tabcolsep{2pt}
\centering\footnotesize
\caption{Key \textsc{TreeFrog} parameters}
\label{tab:treefrog:params}
\begin{tabular}{@{\extracolsep{\fill}}l|l|c|p{0.65\textwidth}}
\hline
\hline
    & Name & Default Value & Comments\\
    \hline
    General
    & $\mathcal{M}_{\rm lim}$ & 0.05 
        & Minimal merit a match must have to be considered a primary link. \\
    \hline
    Descendant
    & $a$ & blah 
        & blah .\\
    \hline
    Progenitor
    & $a$ & blah 
        & blah .\\
\end{tabular}
\end{table*}
\end{comment}

\section{Results}
\label{sec:results:treefrog}
Here we present how well trees are built. As input we primarily use a small cosmological N-Body simulation consisting of $512^3$ particles \cite[from the SURFS suite][]{elahi2018a}. Simulation details are presented in \Tableref{tab:sims}. 
\begin{table}
\setlength\tabcolsep{2pt}
\centering\footnotesize
\caption{Simulation parameters}
\begin{tabular}{@{\extracolsep{\fill}}l|cccc}
\hline
\hline
    Name & Box size & Number of & Particle Mass & Softening \\
    & & Particles & &Length\\
    & $L_{\rm box}$ [$\Mpch$] & $N_p$ & $m_p$ [$\Msunh$] &  $\epsilon$ [$\kpch$] \\
\hline
    L40N512     & $40$  & $512^3$   & $4.13\times10^7$ & 2.6 \\
    L210N1536   & $210$  & $1536^3$ & $2.21\times10^8$ & 4.5 
\end{tabular}
\label{tab:sims}
\end{table}

\par 
We focus on trees produced using two input halo catalogs produced using \textsc{VELOCIraptor}: a simple 3DFOF (3D configuration space Friends-of-friends) catalogue that does not contain subhalos; a 6DFOF (phase-space) halo catalogue; and a full (sub)halo catalogue using fiducial parameters for \textsc{VELOCIraptor}. Details of how \textsc{VELOCIraptor} identifies (sub)halos are presented in \cite{velociraptorpaper}. Here we summarise: the code is a phase-space (sub)halo finder that identifies structures in a two-step process: it identifies field halos using either a 3DFOF algorithm or a 6DFOF algorithm; and then identifies substructures for each halo by linking dynamically distinct particles using a phase-space FOF algorithm and searching for major merger remnants using an iterative 6DFOF. 

\par
\begin{figure*}
    \centering
    \includegraphics[width=0.95\textwidth,trim=0.cm 0.cm 0.cm 0.cm, clip=true]{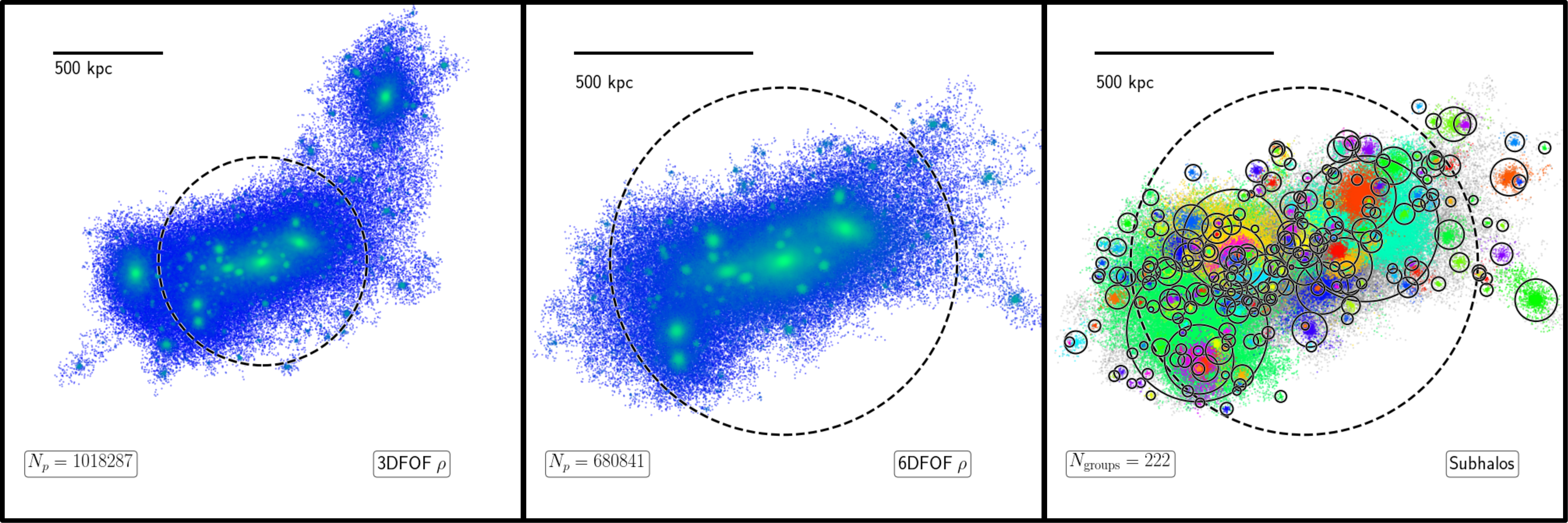}
    \caption{{\bf Example halo:} We show a 3DFOF halo (left), a 6DFOF halo (middle) and the substructure within the 6DFOF halo (right). For each halo we show $R_{\Delta\rho_c}$ by a dashed black circle. In the first two columns, particles are colour-coded according to the 3D density going from blue to green in increasing density. In the right panel, particles are colour-coded by the group to which they belong. We also draw solid circles for each subhalo showing $R_{\Delta\rho_c}$. We show a ruler in each panel. }
    \label{fig:haloexample}
\end{figure*}

 
We show examples from our 3 halo catalogs in \Figref{fig:haloexample}. The 3DFOF halo extracted from our L40N512 simulation at $z=0$ is composed of $\approx10^{6}$ particles with a FOF mass of $4.2\times10^{14}\Msunh$ and an overdensity virial mass of  $M_{\Delta\rho_c}=2.7\times10^{14}\Msunh$, where $M_{\Delta\rho_c}=4\pi\Delta\rho_c R_{\Delta\rho_c}/3$, $\rho_c$ is the critical density, and $R_{\Delta\rho_c}$ is the radius enclosing an average density of $\Delta\rho_c$, where $\Delta=200$, commonly referred to as the virial mass. This 3DFOF object was identified using the standard 3DFOF linking length of $\ell_x=0.2 L_{\rm box}/N_p$, where $L_{\rm box}/N_p$ is the inter-particle spacing. This 3DFOF halo consists of several large density peaks, some of which lie outside the virial radius centred on the largest density peak. The 6DFOF halo is the largest object of the initial 3DFOF candidate and the density peaks that were outside the virial radius of the 3DFOF are now considered separate 6DFOF halos. The 6DFOF halo contains at least 4 large density peaks and numerous smaller ones, speaking to a rich merger history, with several major mergers in the recent past and likely several mergers in the near future. At $z=0$, this object contains 222 substructures (including the host halo), 3 of which contain $21\%$ of the initial host 6DFOF halo's mass. We refer to this halo as our fiducial case as this object has a complex merger history, undergoing a quintuple merger. 

\par 
The trees are built using 200 snapshots spaced evenly in $\log a$, where $a$ is the scale factor, starting at $a_i=0.04$ and ending at $a_f=1$. The cadence is such that at late times the temporal spacing between snapshots is $\sim250$~Myr. We produce several trees for each halo catalogue, varying the merit function and the number of snapshots searched for links. We focus on the descendant based tree as walking forwards in time provides a natural method to correct trees, namely that an object should have a primary descendant. If an object lacks a primary descendant (or any viable descendant), further snapshots can be searched till a primary descendant is found. In contrast, progenitor trees are only corrected for objects that lack a progenitor.   

\par 
For this analysis, we also make use of \textsc{Merger Tree Dendogram} \cite[][]{poulton2018a}. These dendograms capture the mass accretion history of (sub)halos and their orbital evolution. Ideally the mass accretion history of a halo should be smooth, increasing with time, whereas subhalos should slowly shrink, losing most the their mass near peri-centric passage. 

\subsection{Individual Halo}
\label{sec:results:treefrog:halo}
\begin{figure*}
    \centering
    \includegraphics[width=0.8\textwidth,trim=0.cm 2.cm 0.0cm 0.0cm, clip=true]{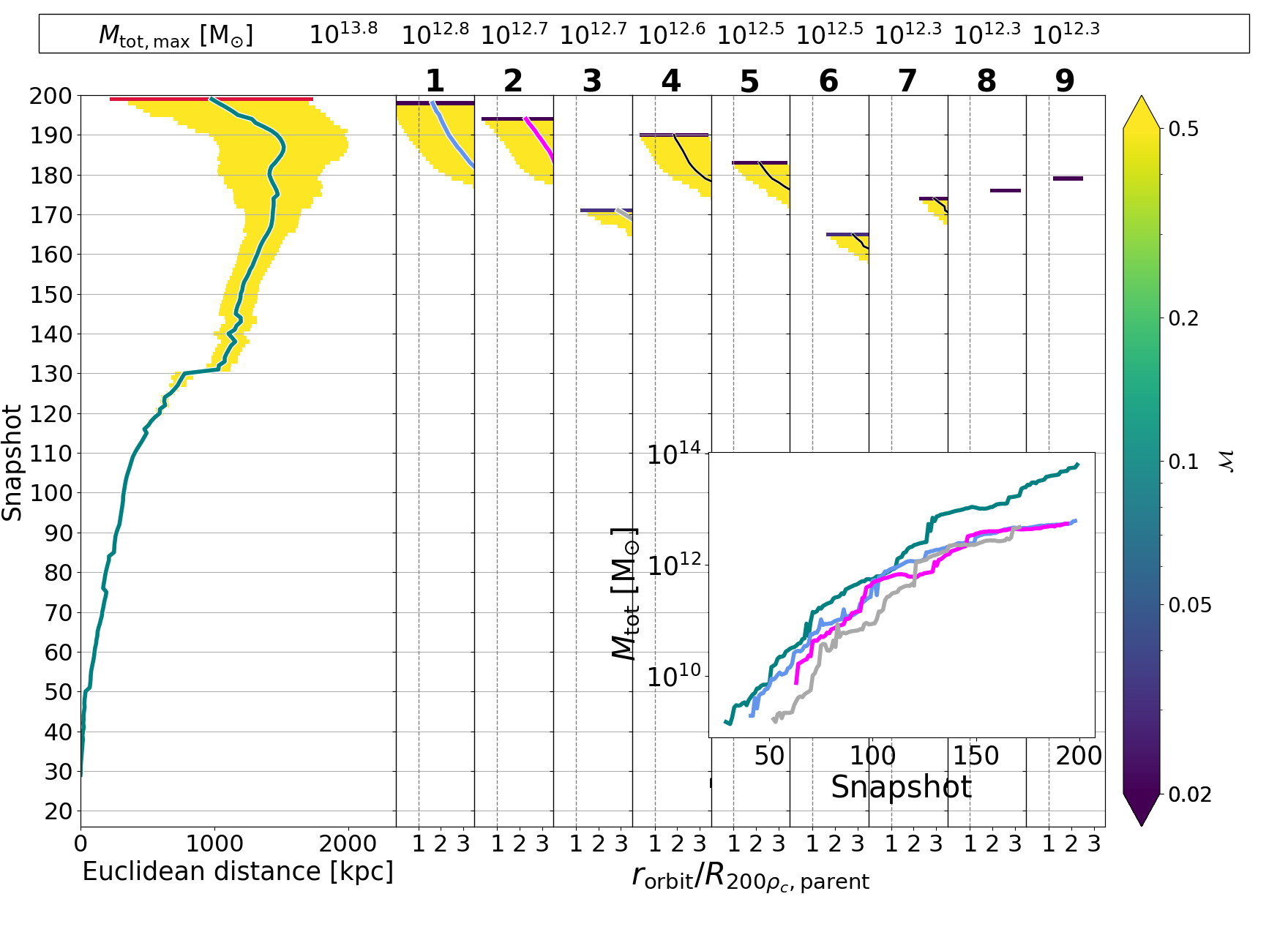}
    \caption{{\bf 3DFOF Example Dendograms:} We plot the mass accretion history and motion of the 3DFOF halo shown in \Figref{fig:haloexample}, along with the relative radial position of a sample of large sub-branches and interacting branches. The first sub-panel on the left shows the motion and mass accretion history of the main branch. Subsequent sub-panels show the motion relative to the main branch scaled by the main branch's virial radius out to 3.5 virial radii. The size of markers indicate the mass of the object, with the size of the markers in the sub-panel increased by a factor of 5 relative to the main branch so as to make their mass evolution more visible. Colours indicate the merit of a match between a halo and its descendant/progenitor depending on the direction of the tree construction. Here we show a descendant tree and use the merit given by \Eqref{eqn:merit} but we do not use the most bound fraction of particles, nor link across multiple snapshots. The range of the colour bar is chosen to emphasise the transition about the nominal acceptable merit of $0.1$. The inset shows the mass evolution of the three largest objects and the main branch. We also show the virial radius in the sub-branch panels by a dashed vertical lines. Any objects that remain an independent object at the last snapshot are marked in red. Note that here, halos show little variation in merit till they merge. Large variations along a branch are seen in other trees.}
    \label{fig:dendo3dfof}
\end{figure*}

\begin{figure*}
    \centering
    \includegraphics[width=0.8\textwidth,trim=0.cm 2.cm 0.0cm 0.0cm, clip=true]{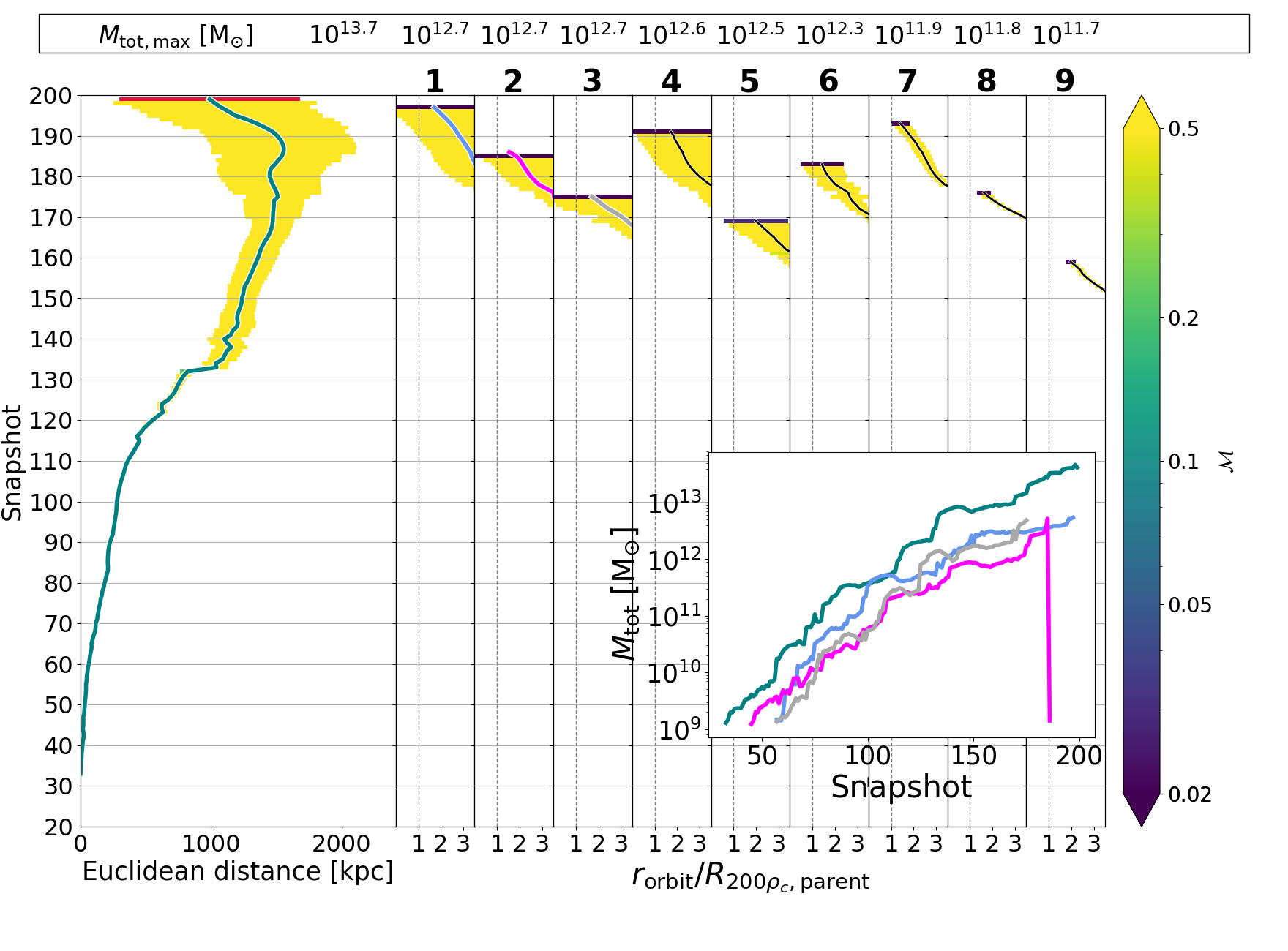}
    \caption{{\bf 6DFOF Example Dendogram:} Similar to \Figref{fig:dendo3dfof} but where we use a 6DFOF input catalogue. Here we again use the merit given by \Eqref{eqn:merit} but we do not use the most bound fraction of particles, no link across multiple snapshots. Using \Eqref{eqn:nshared} does not affect the tree. }
    \label{fig:dendo6dfof}
\end{figure*}

\begin{figure*}
    \centering
    \includegraphics[width=0.8\textwidth,trim=0.cm 2.cm 0.0cm 0.0cm, clip=true]{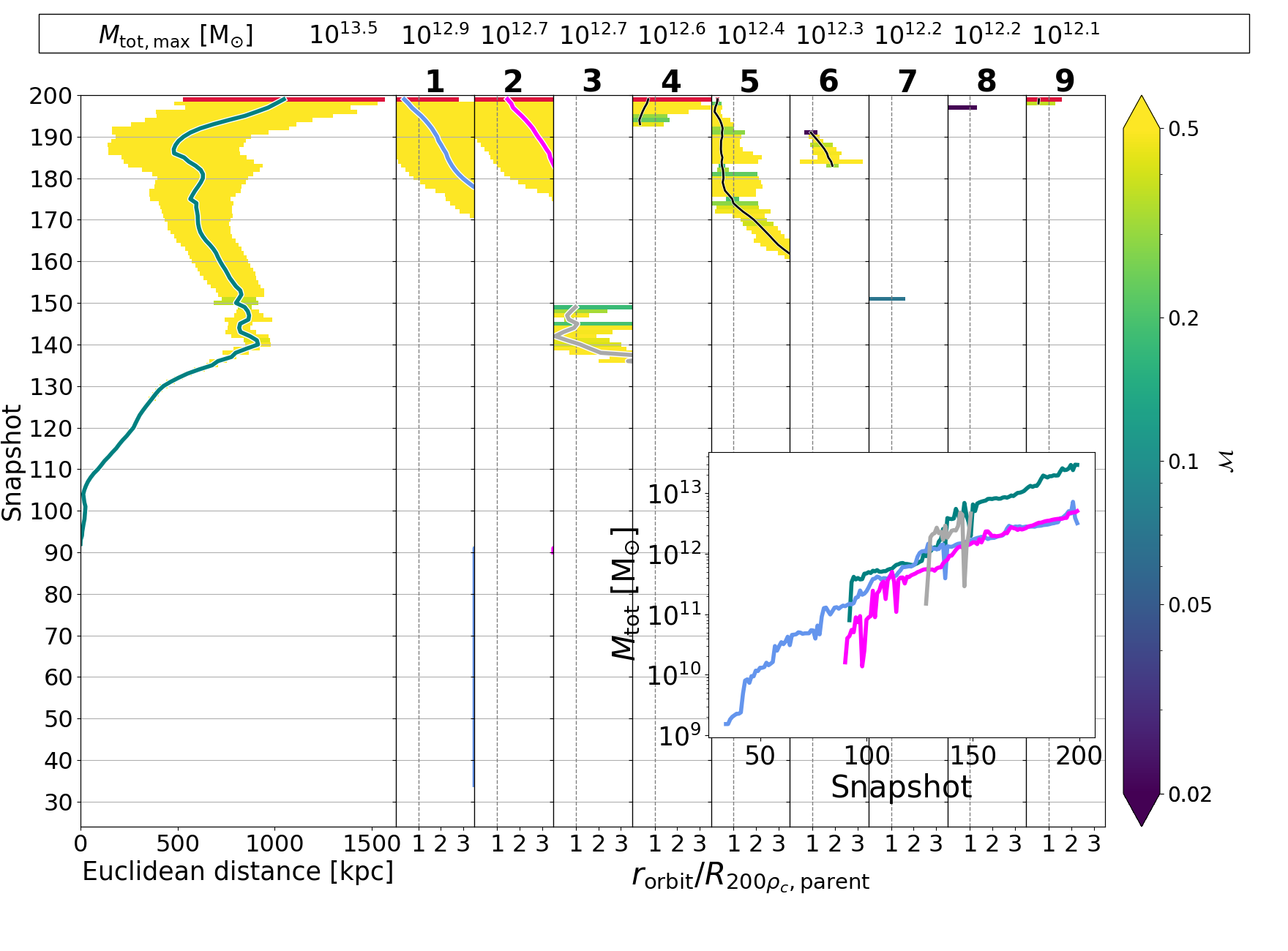}
    \caption{{\bf 6DFOF+Subhalos+Mergers Example Dendogram:} Similar to \Figref{fig:dendo3dfof} but for a halo catalogue that contains both halos and substructure. Note that circles in the bottom sub-panels indicate that this branch has at one point hosted the main branch as a subhalo. Here we again use the merit given by \Eqref{eqn:merit} but we do not use the most bound fraction of particles, no link across multiple snapshots. Using \Eqref{eqn:nshared} in this case does affect the tree, producing smoother mass evolution but introducing kinks in the orbits of objects. Here, variations in merit are seen along individual branches.}
    \label{fig:dendo6dfofsubhalot1}
\end{figure*}

\begin{figure*}
    \centering
    \includegraphics[width=0.7\textwidth,trim=0.cm 2.cm 0.0cm 0.0cm, clip=true]{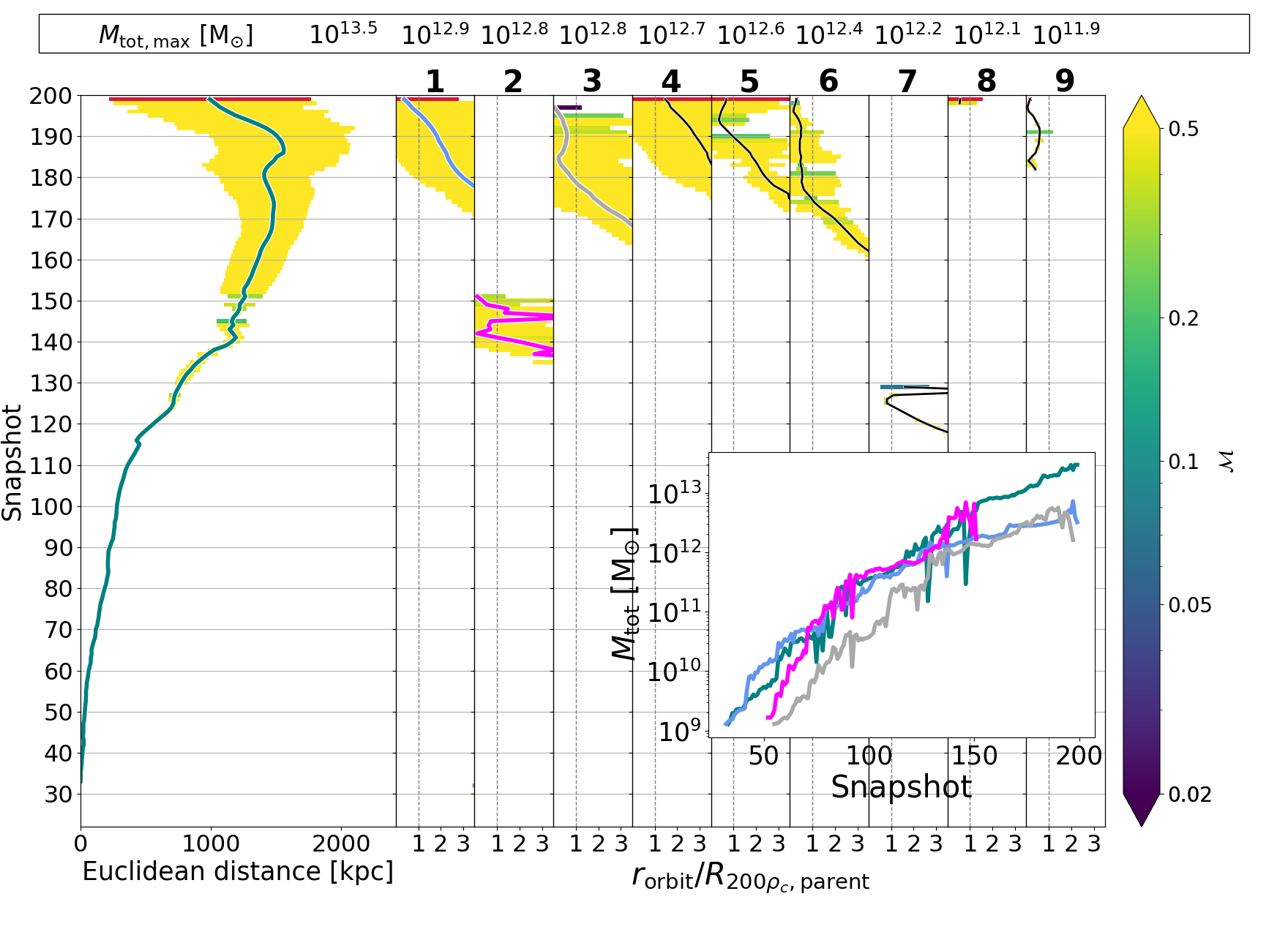}
    \caption{{\bf 6DFOF+Subhalos+Mergers with Corrections:} Similar to \Figref{fig:dendo6dfofsubhalot1} but where we use the merit given by \Eqref{eqn:merit}, use a fraction of the most bound particles, link across multiple snapshots, and apply corrections so as to minimise the number of branch-swapping, flip-flopping and truncation events.}
    \label{fig:dendo6dfofsubhalot4core}
\end{figure*}

We examine the reconstructed merger history of the fiducial halo presented in \Figref{fig:haloexample} in our three halo catalogues using dendograms in figures \ref{fig:dendo3dfof}-\ref{fig:dendo6dfofsubhalot4core}. These figures present the mass and orbital evolution of branches of a tree and any objects the main branch may have interacted with (see Figure 4 from \cite{poulton2018a} which describes in detail the information the dendogram tries to capture). We start with the simplest halo catalogue, the 3DFOF one and build a descendant tree using a single snapshot and a simple merit function to identify links and proceed to add corrections to the tree and complexity to the input (sub)halo catalogue. 

\par 
We present in \Figref{fig:dendo3dfof} the dendogram from the tree built on a simple FOF catalogue with the simplest merit function, \Eqref{eqn:nshared}, using all particles to calculate merits. This dendogram shows the mass accretion history and motion relative to the first progenitor of the main branch, along with the relative radial position of a sample of large sub-branches and interacting branches. We also show the merit between matches via the colour and highlight the mass accretion history of the 4 largest branches in the inset. 

\par 
The figure shows that the FOF halo has a simple mass accretion history.  The main branch halo continuously grows in mass, absorbing smaller halos (sub-branches). There are several kinks it the main branch's motion. These occur during major mergers, where the centre-of-mass can shift, moving to the density peak corresponding to the other halo. The merit of the main branch remains close to unity till the last snapshot, where by construction $\mathcal{M}=0$ as there are no descendants. Using the more complex ranking merit scheme given by \Eqref{eqn:merit} leaves the tree generally unchanged as the input catalogue is simple.

\par 
The sub-branches here merge well outside the main branch's virial radius, a natural outcome of a 3DFOF catalogue. The sub-branches typically have merits close to unity, till they merge with the main branch, where the merit becomes very low (typically $\lesssim10^{-2}$, the exact value depending on the merit function). The sub-branch mass evolution is generally smooth, although at least two sub-branches, number 8 \& 9 are truncated. These objects are actually descendants of sub-branch 7. This object leaves enters and momentarily re-emerges the FOF envelop of the main branch twice. Since we do not allow for halo fragmentation, these objects are left stranded in the tree. These truncation events can be fixed by searching for descendants across several snapshots and using the full merit function given in \Eqref{eqn:merit}. Using all particles to calculate the simple merit in \Eqref{eqn:nshared} incorrectly links these stranded halos to small halos with a very low merit, i.e., branch swapping events. 

\par 
Using a 6DFOF input catalogue corrects some of the issues present in the original tree, indicating how the performance of a tree depends on the input catalogue as seen in \Figref{fig:dendo6dfof}. Objects now merge later, there are fewer truncated large sub-branches but flip-flopping events are still present. There are instances of the large halos linking to small halos, giving rise to the significant drops in mass (see sub-branch 2). Despite a few sub-branches behaving poorly, 3DFOF/6DFOF trees are relatively stable, with the critical issue lying with the misleading physics implied by this tree. Objects should persist till well inside the virial radius. This either requires tracking of FOF particles using codes like \textsc{hbt+} \cite[]{han2018a} or identifying substructure\footnote{The extra complexity introduced by substructure (both in identifying it and determining optimal branches) and the relative simplicity of 3DFOF mergers trees is the motivation behind codes like \textsc{hbt+}, which takes as input the particles in 3DFOF halos and parses them through an unbinding routine to follow their evolution, building a substructure hierarchy and a halo merger tree at the same time. The drawback is that the input 3DFOF catalogue must have high enough cadence to capture the formation of FOF halos and the code requires full particle information across cosmic time.}.

\par 
Figure \ref{fig:dendo6dfofsubhalot1} shows how adding the identification of substructure significantly complicates the process of tree building. This dendogram now contains interacting branches, aka subhalos, as well as the main branch and sub-branches. We can see objects merging well within the virial radius of the main halo. The obvious issues in this halo's reconstructed history are: the main branch starts abruptly and there are several large objects left stranded in the tree with no progenitor. In some cases, the truncation arises from the fact that the halo finder loses track of an object for at least one snapshot. A more subtle issue present is the change in the motion of the main branch. The distance plotted here is the relative comoving distance from the position of the first progenitor. A change in the motion is suggestive of a branch swapping event earlier in the object's history. 

\par
The typical cause of these issues are major mergers. The basic assumption underlying \textsc{TreeFrog} and many tree builders is that particles orbit an individual object and thus can be used to trace the evolution of object. Most of the time a halo grows by the smooth accretion of material or the tidal disruption of smaller objects and the vast majority of particles in the environment principally orbit the potential well defining the main halo. However, the orbits of particles during major mergers are complex. Some particles are ejected from the system altogether, some have orbits that swap the potential well they are orbiting and others orbit both potential wells. The fraction of particles quickly evolving orbits steadily increase with time, starting with the loosely bound particles and progressing to increasingly bound particles as the objects coalesce. Therefore, using all the particles can give rise to flip-flopping, branch swapping and even truncation, clearly seen in \Figref{fig:dendo6dfofsubhalot1}. 

\par
These problems are fixed by: 
\begin{itemize}
    \item Searching for links across multiple snapshots.
    \item Using the most bound particles or ranking particles by how well bound they are to determine the quality of the match. 
    \item Correcting objects with no primary progenitors.  
\end{itemize}
Figure \ref{fig:dendo6dfofsubhalot4core} shows the resulting dendogram once multiple snapshots are searched, here 4 snapshots corresponding to $\sim1$~Gyr or $\sim1$~free-fall dynamical time, $\tau\propto\sqrt{3\pi/32 G\rho}, \rho=200\rho_c$. We also use a fraction of the most bound particles are used, here $0.4$ to calculate merits\footnote{The exact fraction depends on the (sub)halo finder used. Configuration-space (sub)halo finders will artificially shrink large subhalos as they fall to the centre, whereas phase-space finders might overestimate the mass assigned to the infalling object as the object is dynamically heated. For phase-space finders like \textsc{VELOCIraptor}, we find using $\lesssim50\%$ minimises branch swapping events.}, and we correct for missing progenitors/branch swapping events across multiple snapshots. Large subhalos that previously sprang into existence inside the virial radius now are connected. The motion of the main branch is now in better agreement with the original main branch motion seen in \Figref{fig:dendo3dfof}.

\begin{comment}
\par 
These dendograms also show how well \textsc{VELOCIraptor} tracks objects even as they enter deep within their host halo. Sub-branches 4-6 do not show artificial mass loss as they approach pericentre. Only after pericentric passage to objects start to shrink, with large objects having small apocentres due to dynamical friction. The result is smooth mass evolution for both subhalos and halos \cite[not presented here for brevity, see][]{elahi2018a}. However, the recovery is not perfect. Branches 7 and 8 show linear fluctuations in mass along with gaps in the branch. These two objects happen to be subhalos of sub-branches 4 and 5 respectively. There is also a unconnected large object, sub-branch 8, and a small subhalo that terminates early, sub-branch 9. Such missing links are present in all halo catalogs and either are fixed by increasing the number of snapshots searched for links and would benefit from using a halo ghosting tool (like {\sc WhereWolf}, Poulton et al., in prep).
\end{comment}

\subsection{Tree statistics}
\label{sec:results:treefrog:pop}
\begin{figure}
    \centering
    \includegraphics[width=0.49\textwidth,trim=0.5cm 0.cm 0.5cm 1.25cm, clip=true]{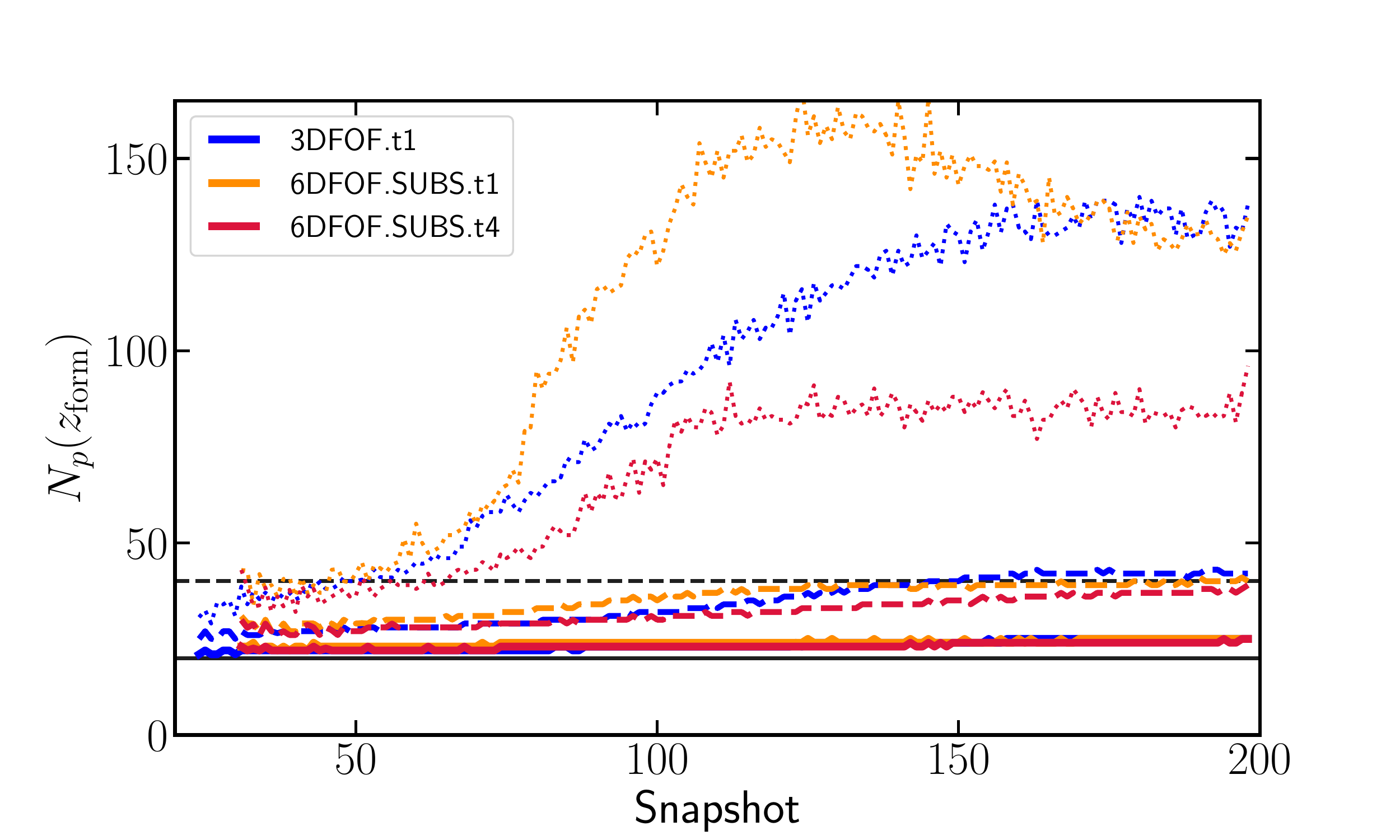}
    \caption{{\bf Particle number at formation:} We plot the particle number at which objects form, that is are identified in the tree as having no progenitor. We show the median, $84\%$ quantiles, and $97.5\%$ quantiles as thick solid, thick dashed, and thin dashed lines respectively. As lower quantiles are similar in all trees and is close to the particle limit at which halos are identified, we do not plot them for clarity. We limit our analysis to snapshots with at least 100 halos. We also show the particle limit at which halos are identified, $N_p=20$, by a solid black line as well as twice this value by a dashed black line.}
    \label{fig:progennpart}
\end{figure}

\begin{figure}
    \centering
    \includegraphics[width=0.49\textwidth,trim=0.5cm 2.0cm 0.5cm 3.cm, clip=true]{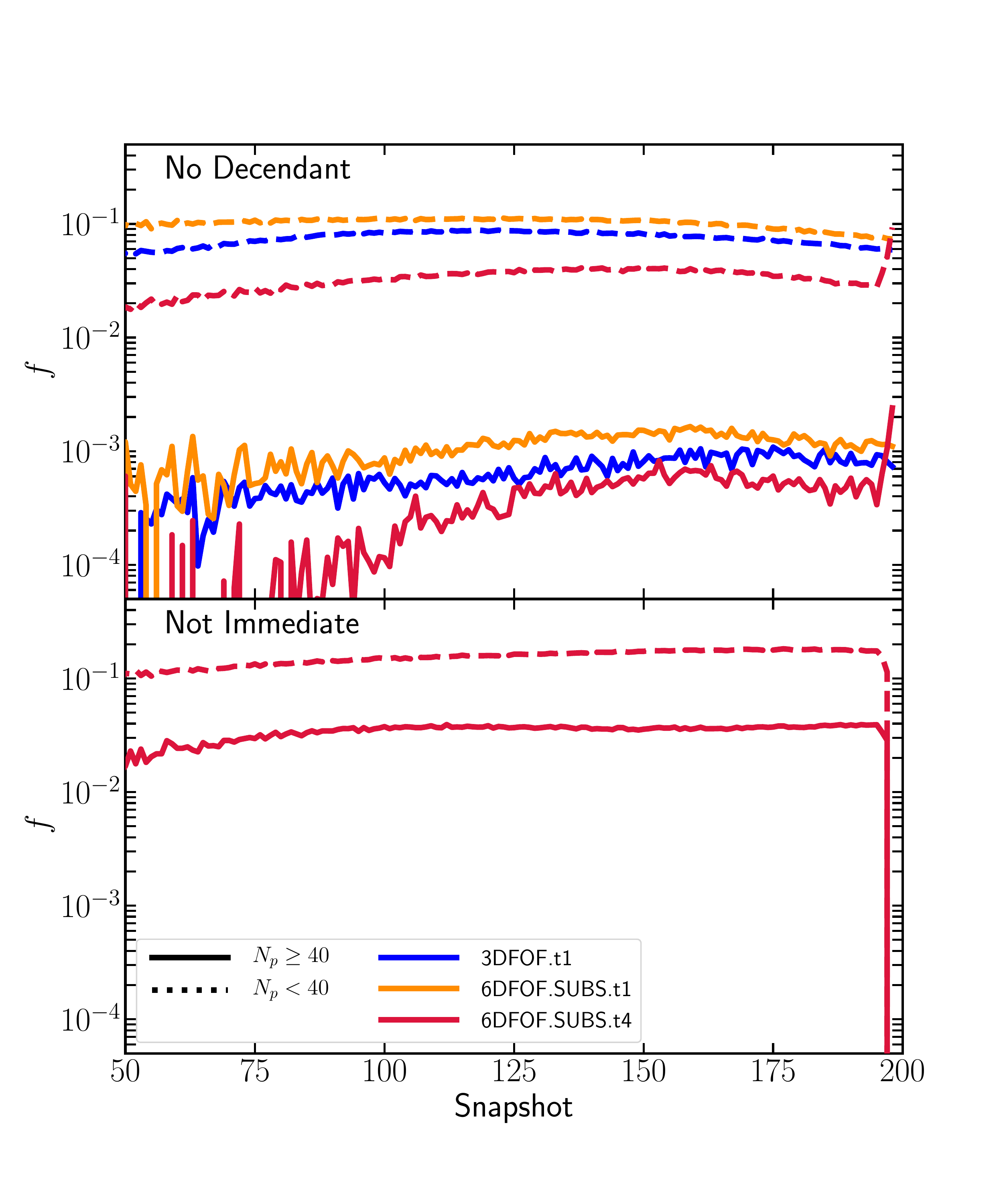}
    \caption{{\bf Non-Ideal Descendant Fraction:} We plot the fraction of objects with either no descendant (top) or a descendant found several snapshots later (bottom) for objects composed of twice the particle limit and those containing fewer particles. We limit our analysis to snapshots with at least 100 halos.}
    \label{fig:fracdescen}
\end{figure}

We now turn to the overall statistics of the tree. Ideally, objects form when composed of few particles and once formed should always have a descendant. Yet poorly resolved objects can evaporate and be left without a descendant, particularly if the time between snapshots is short. We examine the statistics of when objects form and the fraction of objects without a descendant in figures \ref{fig:progennpart} \& \ref{fig:fracdescen} respectively, focusing on the 3DFOF tree, and the 6DFOF+substructure tree using a single snapshot to identify links and 6DFOF+substructure tree built using 4 snapshots. We argue that these statistics are more informative than the common practice to examine branch lengths \cite[e.g.][]{srisawat2013}, i.e., the number of snapshots a main branch exists for, as the length of a main branch depends sensitively on the cadence used in producing the tree. 

\par 
Clearly the median formation particle number in \Figref{fig:progennpart} is very close to the 20 particle limit used to identify structures for all three trees, showing little evolution with $50\%$ of all newly formed objects being composed of $\leq25$ particles. The upper $84\%$ quantiles do show some dependence on cosmic time, increasing from $25$ particles up to close to twice the 20 particle limit. This increase in $N_p(z_{\rm form})$ with cosmic time partly due to the larger physical time between snapshots at late times which allows halos that lie below the 20 particle threshold to grow more. However, the fact that both upper quantiles decrease when using more snapshots to identify links in the tree (going from 6DFOF.SUBS.t1 to 6DFOF.SUB.t4) indicates that this is not the sole reason. Mergers between poorly resolved halos can cause breaks in the tree as one of the halos is lost for one (or more) snapshots before re-emerging, the result being a halo with an artificially inflated $N_p(z_{\rm form})$. The reduction in the upper $97.5\%$ quantile from objects composed of $\gtrsim100$ particles to $\sim80$ when going from using a single snapshot to using 4 snapshots is clearly evidence of truncation. Using multiple snapshots ensures that vast majority of objects form close to the particle limit and even the largest, newly formed objects in the tree are composed of $\lesssim100$ particles at all times. 

\par 
In figure \ref{fig:fracdescen}, we see that the fraction of objects lacking descendants remains roughly constant across most of cosmic time for all trees, decreasing slightly at earlier times. We note that for objects composed of twice the particle limit used in the halo catalogue, this fraction is small $\lesssim10^{-3}$. As we go from the easier problem of following FOF halos to following 6DFOF halos and their substructure we find a slight increase in the fraction if a single snapshot is used. In general, the fraction with no descendants is small and this population is dominated by poorly resolved objects, with $99\%$ composed of $\lesssim40$ particles. Approximately $10\%$ of very poorly resolved objects composed of $<40$ particles have no descendants for trees built using a single snapshot, with the number increasing slightly when using 6DFOF halos. 

\par
Searching multiple snapshots reduces this fraction, dropping it by a factor of 2 for objects composed of $\geq40$ particles, and reducing the amount for objects composed of $<40$ particles to a more reasonable $2\%$. This fraction increases by a factor of 4 in the last 4 snapshots where the number of snapshots used to correct the tree begin to drop, rising from $5\times10^{-4}$ to $2\times10{-3}$. This increase demonstrates the need for searching multiple snapshots and running simulations past the last desired redshift to correct the catalogue at these late times, though the exact number of snapshots depends on the cadence of the input catalogue. For most snapshots, roughly $4\%$ of objects composed of $\geq40$ particles have descendants found more than a single snapshot in the future. 

\par
\begin{figure}
    \centering
    \includegraphics[width=0.49\textwidth,trim=0.5cm 2.0cm 0.5cm 3.0cm, clip=true]{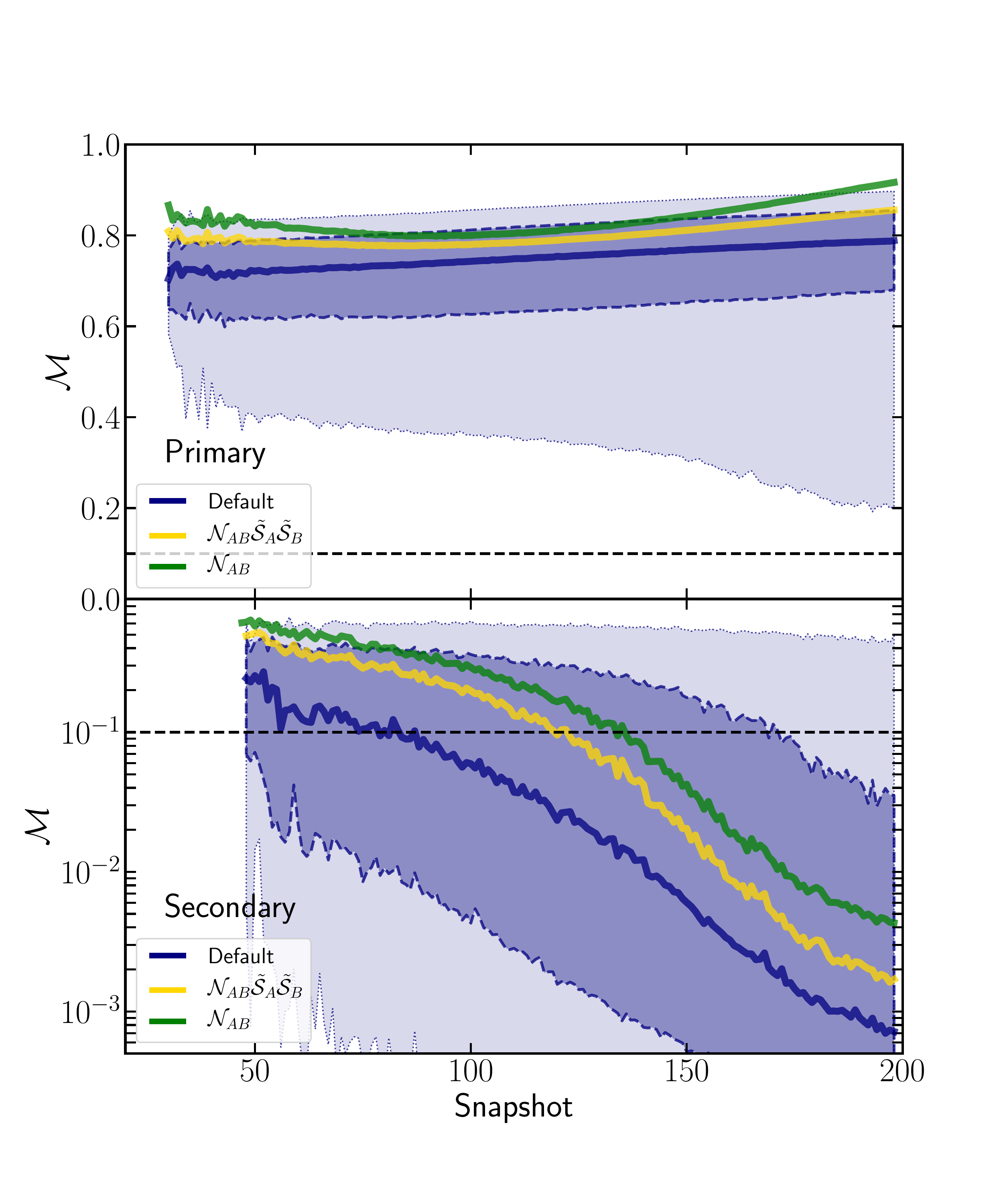}
    \caption{{\bf Merit statistics:} We plot the merit of primary and secondary matches across cosmic time using a single snapshot to identify links. We show the median for our default merit (\Eqref{eqn:merit} using only the 40\% most bound particles), \Eqref{eqn:merit} using all particles, and \Eqref{eqn:nshared} using all particles. For clarity we only show the $16/84\%$ quantiles and $2.5/97.5\%$ quantiles for the default merit, plotted as a dark shaded region outlined by thick dashed lines and a light shaded region outlined by thin dashed lines respectively. We limit our analysis to snapshots where there are at least 100 primary or secondary links and to halos composed of $\geq40$ particles, twice the halo catalogue particle limit. We also show a dashed black line at a nominal good merit of $\mathcal{M}=10^{-1}$.}
    \label{fig:meritstats}
\end{figure}

 
A tree should be constructed so as to have a clear distinction between the main branch and sub-branches. Several merit functions are in common use to rank matches, separating primary descendant/progenitor links that define the main branch and secondary descendant/progenitor links that define sub-branches which merge with the main branch \cite[see][for a sample]{srisawat2013}. The most common merit function maximises the number of shared particles in some form, sometimes using all particles \cite[see for instance][]{ahf}, sometimes using only the most bound set of particles \cite[see for instance][]{dhalos}. \cite{poole2017a} argued for a merit function that used the particles self-binding ranking (Eq. \ref{eqn:ranking}). \textsc{TreeFrog} can use several merit functions, which we show the results of in \Figref{fig:meritstats}. 

\par 
This figure shows that $50\%$ of primary descendants, the merits are close to 1, regardless of the type of merit function used. For our fiducial merit function, which is a combination of using some fraction of the most bound particles and \Eqref{eqn:merit}, we find primary descendant merits of $\sim0.75\pm0.1$, showing little evolution. Only $\sim1\%$ of the population has $\mathcal{M}\lesssim0.2$ and then only at late times. Secondary descendant merits on the other hand are on average $\lesssim10^{-1}$ and evolve strongly with time, a consequence of large, well resolved halos accreting small (sub)halos through natural bottom-up growth. The ever increasing mass ratios probed at late times causes the median secondary merit to drop. 

\par 
Comparing to other merit functions, we find using all particles increases primary merits, and removing the ranking merit, that is using \Eqref{eqn:nshared}, increases the primary merit further. However, this increase in primary merit values is counterbalanced by a similar increase in secondary merits. Simply using \Eqref{eqn:nshared} increases the median secondary merits by a factor of $\sim3$ relative to the default merit function. Critically, the separation between primaries and secondaries is largest using the default scheme. For the fiducial merit, the distribution in merits only overlaps at the $2\sigma$ level at late times. Using the shared number of particles increases the overlap in the population from $\sim3\% $ to $\sim7\%$ (for a more formal comparison of the distances between primaries and secondaries see \Secref{sec:appendix:meritstats}). These changes argues in favour of using \Eqref{eqn:merit} over \Eqref{eqn:nshared}.

\subsection{Mergers}
\label{sec:results:treefrog:mergers}
\begin{figure}
    \centering
    \includegraphics[width=0.49\textwidth,trim=0.5cm 0.5cm 0.5cm 0.5cm, clip=true]{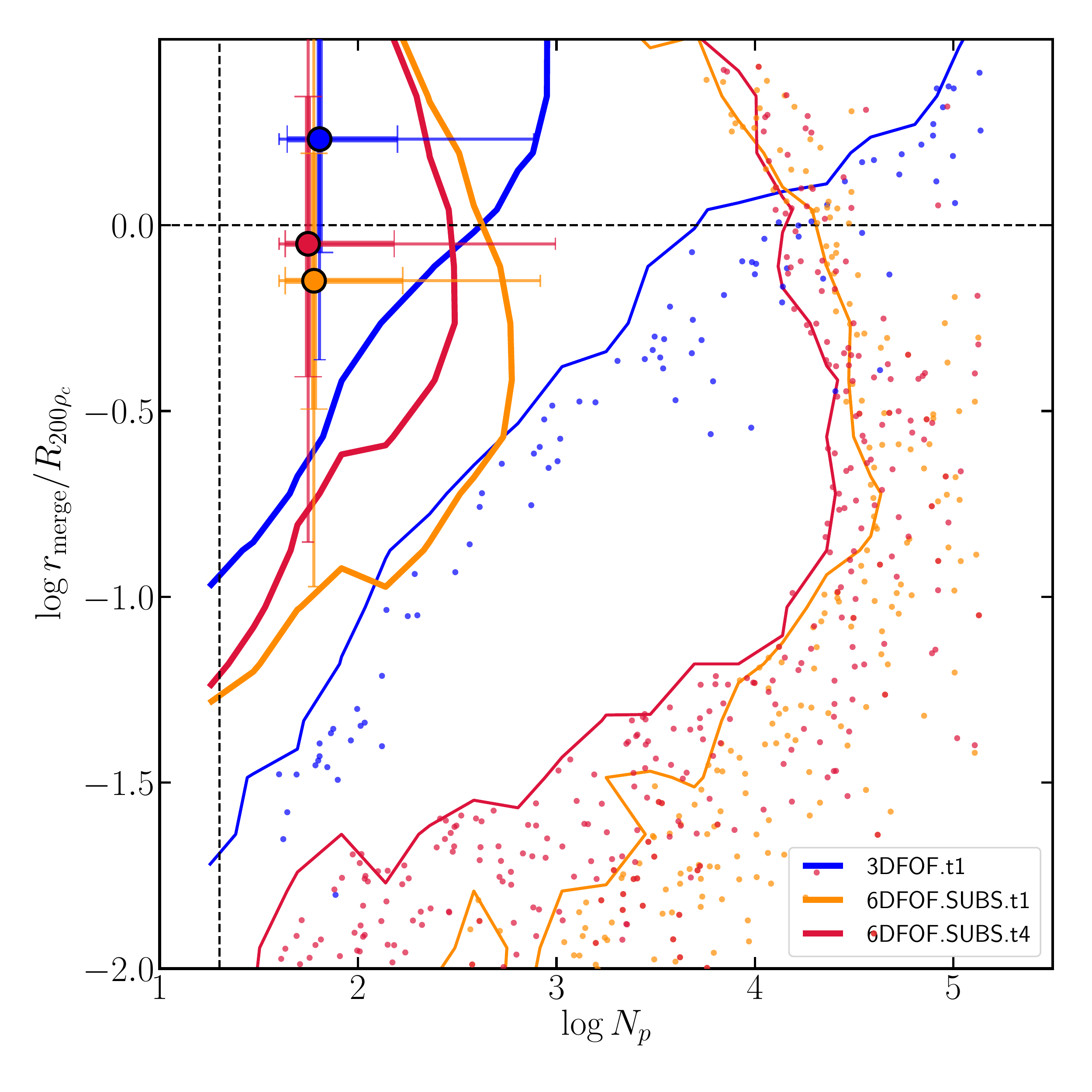}
    \caption{{\bf Merger statistics:} We plot the radius and number of particles at which a (sub)halo mergers with another. Colours indicate the input tree. For each tree, we calculate the median distance and number of particles at which objects merge for objects composed of twice the particle limit used (40) along with the $16,84$ and $2.5,97.5$ quantiles. These are plotted as a circle with thick and thin error bars respectively, coloured by halo merger tree. We also determine the contours that contain $\geq100$ objects and $\geq10$ objects for each tree, denoted by with thick and thin coloured lines. Outliers from the contours are plotted colour coded according to halo merger tree. 
    }
    \label{fig:mergerstats}
\end{figure}

We examine the details of when sub-branches merge with the main branch here. The expectation is that when (sub)halos eventually merge with other (sub)halos as a sub-branch, this should occur well within the virial radius of the accreting (sub)halo. The merger statistics of these trees is shown in \Figref{fig:mergerstats}. Here we have for every (sub)halo across cosmic time identified secondary descendants and noted the relative radial distance the primary and secondary descendant. 

\par 
As expected, trees constructed from 3DFOF catalogues have a majority of objects merging outside the virial radius. The overall distribution of mergers is not only skewed to large radii but larger objects merging at larger radii, which is unphysical as these objects should be less prone to tidal disruption. Of course, this is a natural consequence of using a 3DFOF halo catalogue but is useful for showing an extreme case.

\par 
The merger statistics of the trees built using a full halo+subhalo catalogue show that even for a single snapshot linking most objects merge inside the virial radius and critically, large objects are found to merge well inside accreting halos. Interestingly, using more snapshots to identify descendants shifts the median radius at which objects merge slightly to {\em larger} radii. This is a consequence poorly resolved halos at the outskirts of larger halos that can flicker in and out of the halo catalogue as they drop below/move above the particle limit used in the (sub)halo finder. Without multiple snapshots searched, these objects are left with no descendants (primary or secondary). Searching multiple snapshots links these objects to the larger halo as secondary descendants. 

\par 
An interesting feature in this figure is the presence of large objects that merge at small radii. Naively, the ideal scenario is for objects to merge when composed of few particles deep within another object. However, a natural consequence of major mergers is that objects can phase-mix while still relatively intact. For mini mergers, where the accreted object is far less massive than the accreting object, we expect the smaller object to orbit several times as it is tidally stripped, shrinking to the point at which it becomes completely tidally disrupted. Thus, we should see objects with large accretion masses relative to their accreting host merging at smaller radii, possibly with large masses, compared to mini mergers with very small mass ratios. We note that by accretion, we mean the point at which the object enters the FOF envelope of another halo.

\par
\begin{figure}
    \centering
    \includegraphics[width=0.49\textwidth,trim=0.8cm 3.5cm 1.5cm 1.0cm, clip=true]{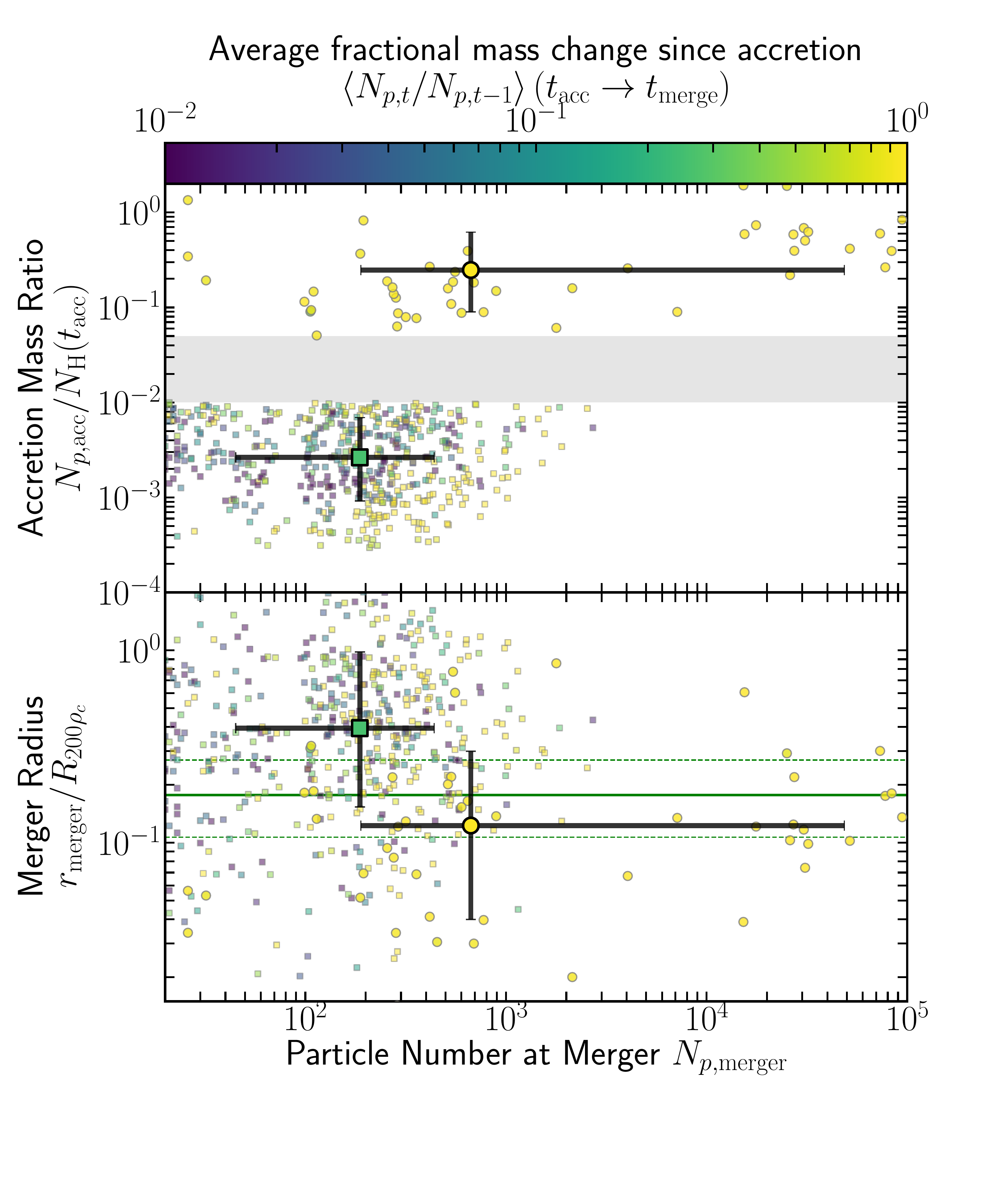}
    \caption{{\bf Minor/Major versus Mini Mergers:} We plot the merger particle number of a sample of objects, exploring the differences between mini and minor/major mergers. Top panel shows the ratio of between the number of particles in the accreted halo and the accreting halo at the time of accretion. Bottom panel shows the radial position where the accreted halo merges with its host. We separate accreted objects into major ($N_{p,{\rm acc}}/N_{\rm H}(t_{\rm acc})\geq5\times10^{-2}$) and minor $N_{p,{\rm acc}}/N_{\rm H}(t_{\rm acc})\leq10^{-2}$ mergers, plotting circles and squares respectively. We also show the median value for each population along with the 16/84 quantiles by large points. All points are colour-coded by the median ratio of the object's current number of particles to that in the previous time step for all snapshots post-accretion, a measure of the mass loss rate.}
    \label{fig:majorvsmini}
\end{figure}

 
To explore this dichotomy, we take a random sample of accretion events that fully merge before $z=0$ with halos composed of $\geq10^5$ particles, splitting the population by the accretion mass ratio. We split objects based on accretion mass ratios into those with ratios of $\leq10^{-2}$ (mini mergers) and those with $\geq5\times10^{-2}$ (containing both minor to major mergers), although the precise split is not critical. The host halo limit and ratio cuts means that minor/major mergers are composed of $\gtrsim5000$ particles at accretion. We should stress that this sample is biased as we are focusing on objects that were accreted and then merger with the host within $\lesssim3$~Gyr. Many objects do not merge with their host halo within this time and are not present in this figure, consequently the mass loss rate of this population is higher than the full population. For each merged object we determine the average mass change from one snapshot to the next since accretion till it mergers. We plot the total population and the median values of accretion mass, particle number at merger, merger radius and mass change in \Figref{fig:majorvsmini}, which has two key features. 

\par
The first feature of note is that on average, so-called minor/major mergers fully merge at smaller radii than mini mergers (objects with very small accretion mass ratios). The former events typically merge inside the scale radius of the host halo (see median values), latter outside. The merger radius does not show a dependence on particle number at accretion for either population. 

\par 
Second, on average mini mergers steadily lose mass as they orbit, on average losing $\sim75\%$ of their mass every 250 Myr (as indicated by the colour). The average fractional mass change does show some dependence on the accretion mass ratio of mini mergers, with small mini mergers ($N_{p,{\rm acc}}/N_{\rm H}(t_{\rm acc})\lesssim 10^{-3}$) show little average change in mass till merging. In contrast, objects with large accretion mass ratios typically do not steadily lose mass once accreted. Instead, they rapidly phase-mix when they enter the central regions of the host halo and typically remain in the central regions due to dynamical friction \cite[][also finds large subhalos are ``trapped'' in the central regions of their host due to dynamical friction using \textsc{hbt+}, a 3DFOF tracker]{han2018a}. These objects can fully merge with the host halo while still close to their accretion mass, such as the sub-branch seen in \Figref{fig:dendo6dfofsubhalot4core}. Others are last identified when composed of a few hundred particles having been accreted when composed of several thousand particles, with most of the mass loss occurring in the last step at which the object was identified. 

\par 
The physical imprint of accretion mass on the dynamics of the merger is reproduced by the tree, though recovering this bimodal distribution requires a phase-space finder that does not artificially shrink halos as they fall inwards. The high mass loss rates of minor/major merger as they phase-mix means that trees should cross match only the most bound particles and ranking particles according to binding energy in order to recover the last inspiral, although using too few particles can give rising to core swapping, were a small subhalo takes over the branch. 

\subsection{Accretion and Merger rates}
\label{sec:results:treefrog:rate}
\begin{figure}
    \centering
    \includegraphics[width=0.49\textwidth,trim=0.cm 5.cm 1.25cm 5.0cm, clip=true]{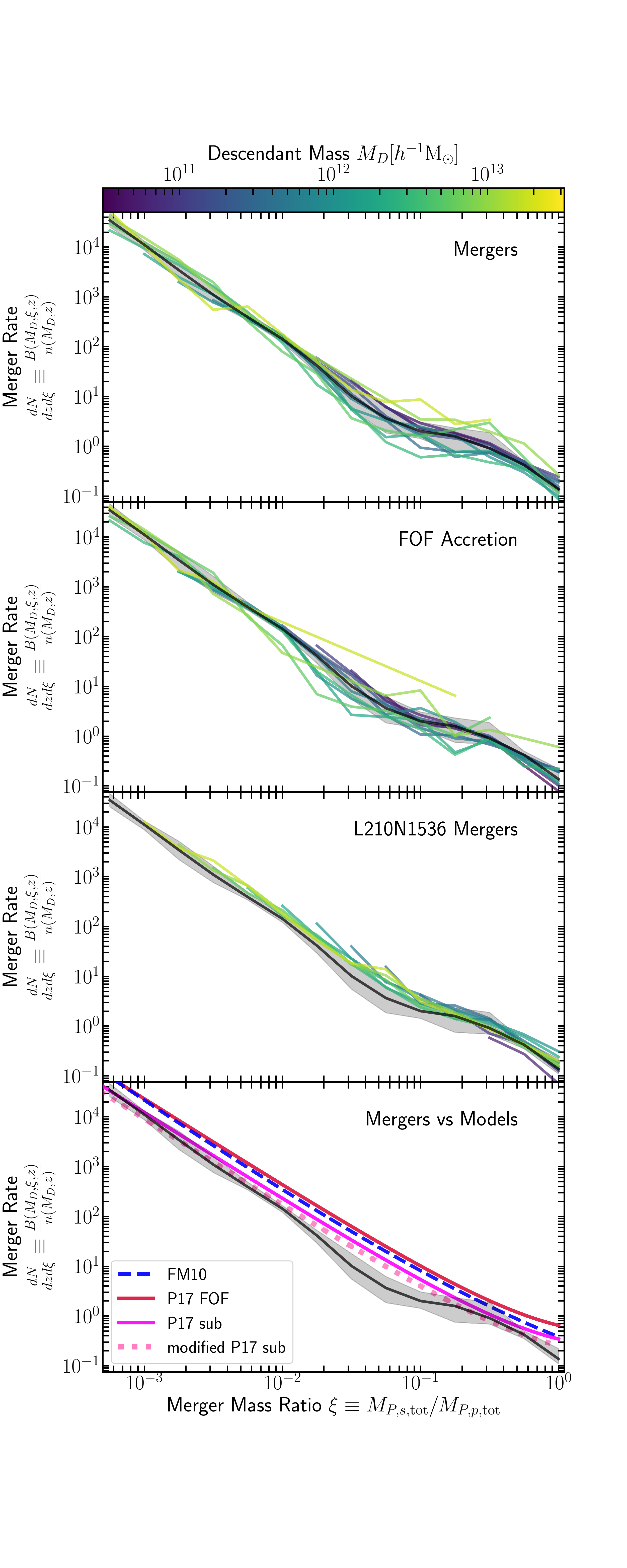}
    \caption{{\bf Mass Accretion Rate:} We plot the rate per halo at which objects merge with a main-branch (solid lines) from $z=[0.5,0]$ for all objects in L40N512 run in the top subpanel labelled 'Mergers'. We also show the median curve and the scatter by a solid black line and a gray shaded region. The next panels compare the median merger rate to accretion rate of FOF halos (second subpanel), comparison to L210N1536 (third subpanel), and a comparison to several models (bottom subpanel). The models shown are the FOF merger rate fit from \cite{fakhouri2010a} (FM2010), the FOF merger rate fit form \cite{poole2017a} (P17 FOF), the \cite{poole2017a} ``corrected substructure accretion rate'' (P17 sub), and a modification of the fit (see text for details).}
    \label{fig:mergerrate}
\end{figure}

We end by examining the merger rates and mass growth via mergers. We calculate the ``{\em mean merger rate per halo}'' expressed in terms of redshift, descendant mass and mass ratio between primary progenitors and secondary progenitors in \Figref{fig:mergerrate}. For every halo with mass $M_D$ having multiple progenitors, we determine the mass ratio between the primary progenitor and secondary progenitors, $\xi\equiv M_{P,s, {\rm tot}}/M_{P,p, {\rm tot}}$, where we use the total exclusive mass associated with the object and bin in mass ratio bins, averaging over a redshift range of $z=0.5$ to $0$. We also calculate the ``{\em mean accretion rate of FOF halos per halo}'', that is we identify all objects that are progenitors or substructures of a descendant halo which were FOF halos at the previous snapshot, that is we define accretion as a FOF halo entering the FOF envelop of another, more massive, FOF halo. This FOF halo can survive as a subhalo or can merge with the accreting FOF halo. This ``{\em FOF accretion rate}'' is analogous to the ``{\em FOF merger rate}'' reported in \cite{fakhouri2008a,fakhouri2010a,genel2010a}, who examined the rate at which FOF halos merge with one another. This is also analogous to the ``{\em corrected substructure merger rate}'' presented in \cite{poole2017a}\footnote{Both accretion and merger are used in the literature, sometimes describing the same physical process, which can be a bit confusing. Here we refer to mergers as instances where an object is found to be a secondary progenitor of another object later in time, that is the object has ceases to exist as an independent object, and is a secondary branch of another object that continues to exist. Accretion events are specifically when a FOF object enters the FOF envelop of another more massive object.}. 

\par
\cite{fakhouri2010a} showed that this FOF merger rate has a nearly universal dependence on $\xi$, with little dependence on descendant mass and redshift, and is characterised by:
\begin{align}
    \frac{dN_{\rm FOF,m}}{dzd\xi} &\equiv \frac{B(M,\xi,z)}{n(M,z)}\notag\\
    &=
    A\left(\frac{M}{10^{12}{\rm M}_\odot}\right)^\alpha(1+z)^\eta
    \left\{\xi^\beta\exp\left[\left(\frac{\xi}{\tilde{\xi}}\right)^\gamma\right]\right\},
    \label{eqn:fofmergerrate}
\end{align}
where $B$ is the number of mergers per unit volume, redshift and mass ratio, $n$ is the number density of halos, and $A,\alpha,\eta,\beta,\tilde{\xi},\gamma$ are all fitting parameters. Fits show $\alpha,\eta$ are small, indicating a weak dependence on descendant mass and redshift. 

\par
We find that both merger and accretion rates have forms similar to the FOF merger rate of \cite{fakhouri2010a}, a roughly power-law dependence on $\xi$ with a flattening at large mass ratios with little descendant mass dependence (as seen by the fact that the coloured lines overlap in the top subpanel). We also find little merger rate differences between our small volume L40N512 run and our larger volume, poorer mass resolution L210N1536 run, indicating a converged merger rate. The differences between our rates and the fits of \cite{fakhouri2010a} (and \citealp{poole2017a}) are that the rate is lower and the high mass ratio rate is flatter. The amplitude is closer to the \cite{poole2017a} ``corrected substructure merger rate'' fit, though this rate is still higher. The reduced rate is a result of defining halos as 6DFOF objects rather than 3DFOF objects \footnote{3DFOF objects can artificially join halos by thin particle bridges, effectively pushing back the accretion time. Using a 6DFOF algorithm removes these particle bridges, moving the accretion time to the point at which the virialized envelops begin to dynamically overlap, increasing the number of halos,  see \cite{velociraptorpaper}}. Modifying the ``corrected substructure merger rate`` fit from \cite{poole2017a} to account for the larger number of 6DFOF objects moves this fit into better agreement with the measured accretion rate. 

\par 
There is little difference between the measured merger and accretion rates despite the fact that a significant amount of time (and evolution) can elapse before an accreted object fully mergers. The mapping from FOF accretion rates to merger rates is non-trivial. Merger time scales depend on the orbit, the tidal mass loss rate, and the initial accretion mass ratio (subhalos with large ratios should experience little tidal mass loss). Subhalos can merge through a variety of channels, some artificial (lost by the (sub)halo finder, numerical evaporation) and some physical (tidal disruption, phase-mixing). At $\xi\approx1$, one might expect a simple delay between accretion and merging as large subhalos will quickly sink to the centre of the host due to dynamical friction, leaving the the functional form unchanged. As one transitions from the regime dominated by dynamical friction to tidal mass loss, near $\xi\sim5\times10^{-2}$, the mapping from accretion to merging becomes more complex and we might expect a change in the functional form. This does appear to be the case, with the largest difference between these rates occurring between $10^{-2}\lesssim\xi\lesssim10^{-1}$. The 
%However, the fact that the merger rate has the same high mass ratio behaviour as the accretion rate, where mass loss should be small and merger time scales short due to dynamical friction shows that \textsc{VELOCIraptor} does not artificially shrink subhalos as they merger and that \textsc{TreeFrog} is able to track mergers. 

\par
The merger rate indicates that halos, on average, experience more minor mergers than major ones but will acquire more mass during major mergers. Based on the modified fit, we expect $10^{12}\Msunh$ halos at $z=0$ to have experienced $\sim10$ minor merger events ($\xi=[10^{-3},5\times10^{-2}]$) from $z=1$ to $z=0$ compared to a single major merger event ($\xi>5\times10^{-2}$), yet halos gain most of this mass in a single major merger, $18\%$ compared to $5\%$. The fit predicts halos should acquire $\approx20\%$ and $\approx79\%$ of their mass integrated over cosmic time through minor and major mergers respectively. In agreement with this prediction, we find $20\pm10\%$ and $31_{-13}^{+46}\%$ are accreted through these two channels\footnote{The prediction ignores ``smooth'' mass accretion whereas halos in our simulation do accrete material not contained within smaller halos, accounting for $\approx46\%$ of a halo's mass growth.}.

\par
\begin{figure}
    \centering
    \includegraphics[width=0.49\textwidth,trim=0.5cm 2.0cm 1.5cm 3.0cm, clip=true]{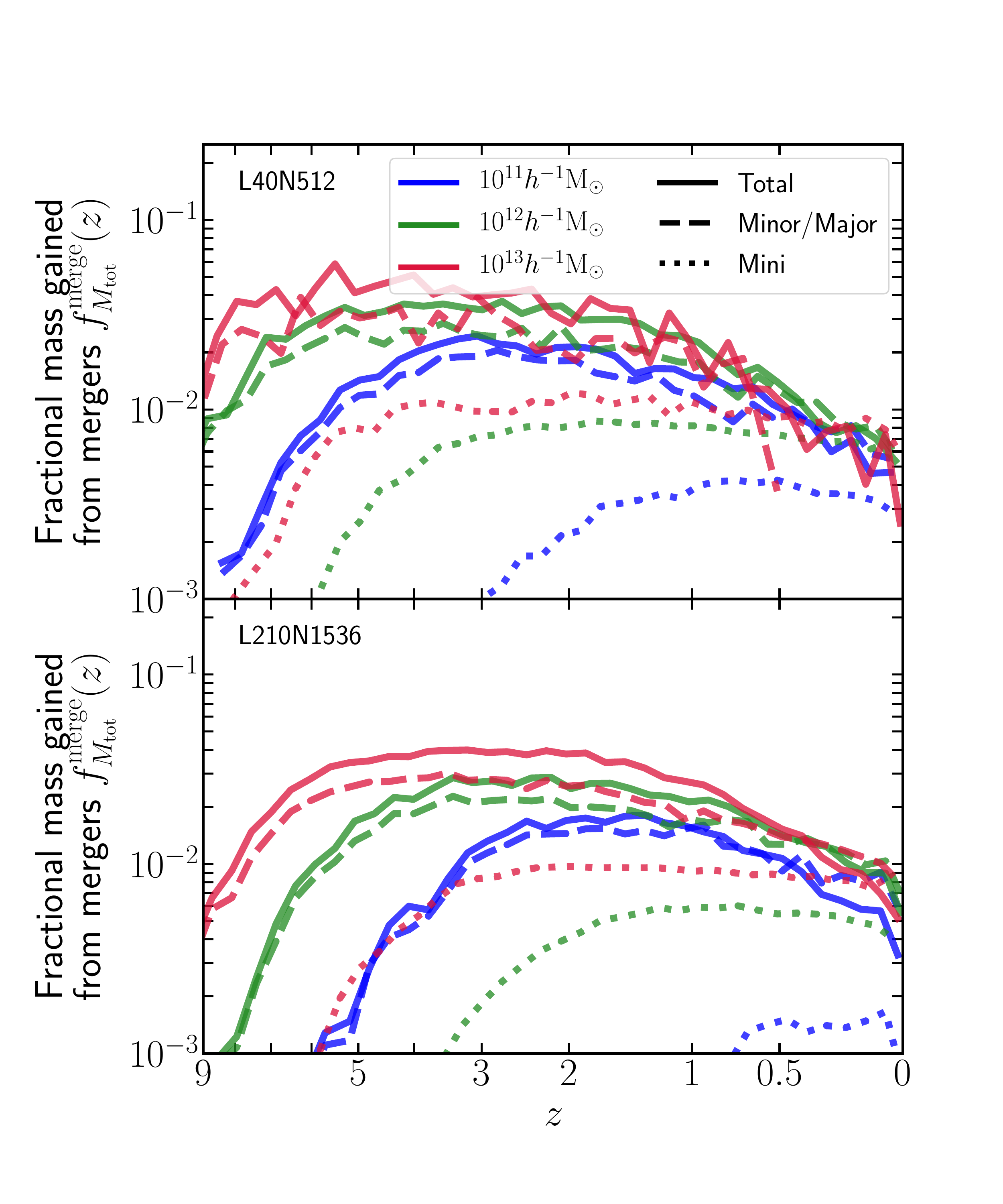}
    \caption{{\bf Mass Growth Through Mergers:} We show the average fraction of mass by which halos grow $f_{M_{\rm tot}{\rm merge}}$, where we use the peak accretion mass of the object that merges to calculate the mass increase in the host halo, as a function of cosmic time along. We also categorise merger events as minor/major and mini mergers based on the accretion mass ratio, with minor/major and mini mergers having accretion mass of $>5\times10^{-2}$ \& $\leq5\times10^{-2}$ respectively. We plot the average for each of these categories for our reference simulation, L40N512 (top), and our larger volume, lower mass resolution simulation, L210N1536 (bottom).}
    \label{fig:mergerchannels}
\end{figure}

 
The total mass accreted in merger events and its evolution is shown in \Figref{fig:mergerchannels}, where we split halos into three different $z=0$ mass bins. At all times, halos principally grow through minor/major mergers, on average gaining $\sim85\%$ of their mass during such events. The overall amount of mass acquired through high mass ratio mergers is greater than that acquired in mini mergers. Initially, it appears objects only grow through minor/major mergers at high redshift, however, this is partially due to finite resolution. 

\par 
The influence of resolution can be seen by comparing results from our reference L40N512 to our larger volume, lower mass resolution simulation, L210N1536. In the lowest mass bin, $10^{11}\Msunh$, halos are composed of $700-7000$ particles in L40N512 compared to $140-1400$ in L210N1536, with the bin dominated by lower mass objects (as a result of the mass function). The total mass gained by $z=0$ is similar for these halos in both simulations, $58^{+31}_{-20}\%$ compared to $54_{-19}^{+28}$, where the uncertainties indicate the halo-to-halo scatter. However, the amount of material accreted through mini mergers in the L210N1536 simulation is significantly reduced. The halos in L210N1536 are not well resolved enough to follow mini mergers and only experience mini mergers at late times, after $z<1$, gaining only $2\%$ of their mass via this channel and only for the largest halos in this mass bin. Improving the mass resolution results in mini-mergers occurring as early as $z\sim3$, with halos gaining $14_{-8}^{+12}\%$ of their mass through mini mergers. This will impact the internal mass distribution of halos as minor/major mergers centrally deposit their mass.

\par 
The mass accretion also shows the imprint of a finite simulation volume, particularly at late times. The largest halos in the smaller volume L40N512 run have a smaller amount of mass acquired through minor/major mergers than similar mass halos in the larger volume L210N1536 run, with few objects experiencing major mergers after $z=0.5$. The total amount of mass acquired in these minor/major mergers is $26_{-15}^{+23}\%$ compared to $36_{-17}^{+32}\%$. The inclusion of large-scale power in L210N1536 gives rise to rarer density peaks, altering the mass accretion rate onto these peaks. \cite{klypin2018a} show that the $z=0$ halo mass function at high masses is suppressed in smaller simulation volumes, particularly at cluster mass scales in agreement with theoretical predictions \cite[also see for instance][for discussion of finite volume effects on power-spectra and the halo mass function]{bagla2006a,warren2006,tinker2010a,schneider2016a,comparat2017a}. Finite volume effects on the mass accretion history has yet to be thoroughly investigated and is beyond the scope of this paper. 

\section{Discussion and Conclusion}
\label{sec:discussion}
We have presented \textsc{TreeFrog}, a code designed to follow the evolution of cosmic structures like halos and subhalos. We have demonstrated that the code tracks (sub)halos across cosmic time, particularly cases that are typically notoriously difficult for such codes, namely major mergers. We summarise key features and results below.  

\par 
\textsc{TreeFrog} is a tree builder code that can take a variety of halo catalogue inputs. At its core, it is a particle correlator, using particle IDs to match halo catalogues. Used in concert with \textsc{VELOCIraptor} (or any halo finder where the input particle lists are arranged in a meaningful order), it uses a combination of merit functions along with a subset of the most bound particles to determine the best matches. We have shown that searching multiple snapshots for candidate descendants based on the combined rank ordering/number of shared particles merit using $\sim50\%$ of the most bound particles well reconstructs the accretion histories of objects with complex interactions (those that experience major mergers and host significant amounts of substructure), even for objects which contain substructure and become subhalos of a larger host.

\par 
The combined merit used by \textsc{TreeFrog} better separates primary progenitors/descendant links from secondary ones, with the primary and secondary merit distributions overlapping at the $2.5\%$ level, unlike the commonly used number of shared particles based merit, where the distributions overlap at the $7\%$ level. Searching multiple snapshots for possible descendants is critical, reducing the number of very poorly resolved objects with artificially truncated lives from $\sim10\%$ to $\sim2\%$. The reduction in truncation events and other tree pathologies advocates the need to run simulations {\em into the future}, past the desired last redshift, a practice also advocated by \cite{poole2017a}.

\par 
The net result is that only a small fraction of objects either start or end life composed of too many particles. Less than $1\%$ of objects begin their lives composed of $\gtrsim100$ particles, above the halo catalogue particle limit of 20. A negligible fraction of objects, $\sim10^{-4}$, composed of $\sim40$ particles have artificially truncated lives, ending with no descendant. Typically, these objects are poorly resolved halos in the process of being tidally disrupted that are falling towards another halo. 

\par
With well built trees, we find a significant difference in the merger behaviour of small subhalos and major merger remnants. Mergers, those objects that are accreted by another halo with accretion mass ratios of $\gtrsim5\times10^{-2}$ fully coalesce or merge at smaller radii that subhalos, those with accretion masses ratios of $\lesssim10^{-2}$, due to the effect of dynamical friction. These objects do not experience significant tidal mass loss, the slow stripping of outer less bound material. Instead, they begin to phase-mix, with the mass assigned to the merger remnant rapidly being assigned to the host halo once they move close to or inside the scale radius of the host halo. 

\par 
We find that the mass accretion history of a host halo is dominated by major mergers. In agreement with previous studies, we find $20\pm10\%$ of the mass is accreted through minor mergers and $31_{-13}^{+46}\%$ through major mergers. The reconstructed merger rate from our low resolution simulation is in agreement with those from \cite{poole2017a} using much high resolution simulations. 

\par 
The general particle correlator nature of \textsc{TreeFrog} means that it has been used to not only construct halo merger trees but void trees \cite[][]{sutter2014a} and even compare halos across simulations with different subgrid physics \cite[e.g.][]{nifty3,nifty5}.

\par
The process of developing \textsc{TreeFrog} has lead to the spin-off of two standalone packages {\sc WhereWolf}, an halo tracking tool for halo merger trees which fills in gaps in the tree and follows objects deemed to have merged; and {\sc OrbWeaver}, a tool to reconstruct orbital evolution, that will be presented in a follow-up paper (Poulton in prep). The former corrects gaps and mergers in the tree by tracking particles belonging to the (sub)halo that has a gap in the tree or has merged, to see if the (sub)halo still exists. The latter reconstructs the orbital evolution of halos. 

\begin{acknowledgements}
RP is supported by a University of Western Australia Scholarship. RC is supported by the SIRF awarded by the University of Western Australia Scholarships Committee, and the Consejo Nacional de Ciencia y Tecnolog\'ia (CONACyT) scholarship No. 438594 and the MERAC Foundation. Parts of this research were conducted by the Australian Research Council Centre of Excellence for All Sky Astrophysics in 3 Dimensions (ASTRO 3D), through project number CE170100013. CL is funded by a Discovery Early Career Researcher Award DE150100618. CL also thanks the MERAC Foundation for a Postdoctoral Research Award. 

\par
The authors contributed to this paper in the following ways: PJE ran simulations and analysed the data, made the plots and wrote the bulk of the paper. PJE is the primary developer of {\sc TreeFrog}. RP \& RT designed and developed various aspects of the code: RP tested the code and motivated the development of the descendant tree; RT developed the compilation infrastructure. RC, CL, CP, \& AR assisted in the design of various aspects of the code. All authors have read and commented on the paper.
\paragraph*{Facilities} Magnus (Pawsey Supercomputing Centre)
\paragraph*{Software} 
\begin{itemize}
    \item \textsc{VELOCIraptor} \href{https://github.com/pelahi/VELOCIraptor-STF}{\url{https://github.com/pelahi/VELOCIraptor-STF}}
    \item \textsc{TreeFrog} \href{https://github.com/pelahi/TreeFrog}{\url{https://github.com/pelahi/TreeFrog}}
    \item \textsc{NBodylib} \href{https://github.com/pelahi/NBodylib}{\url{https://github.com/pelahi/NBodylib}}
    \item \textsc{VELOCIraptor\_Python\_Tools} \href{https://github.com/pelahi/VELOCIraptor_Python_Tools}{\url{https://github.com/pelahi/VELOCIraptor_Python_Tools}}
    \item \textsc{MergerTreeDendograms} \href{https://github.com/rhyspoulton/MergerTree-Dendograms/}{\url{https://github.com/rhyspoulton/MergerTree-Dendograms}}
\end{itemize}
\paragraph*{Additional Software} Python, Matplotlib \cite[][]{matplotlib}, Scipy \cite[][]{scipy}, emcee \cite[][]{emcee}, SciKit \cite[][]{scikit}, Gadget \cite[][]{gadget2}
\end{acknowledgements}

\bibliographystyle{pasa-mnras}
%\bibliography{researchbib}
\bibliography{treefrog.bbl}

\begin{appendix}
\section{Merit Function Comparison}
\label{sec:appendix:meritstats}
\begin{figure}
    \centering
    \includegraphics[width=0.49\textwidth,trim=0.5cm 0.cm 0.5cm 1.0cm, clip=true]{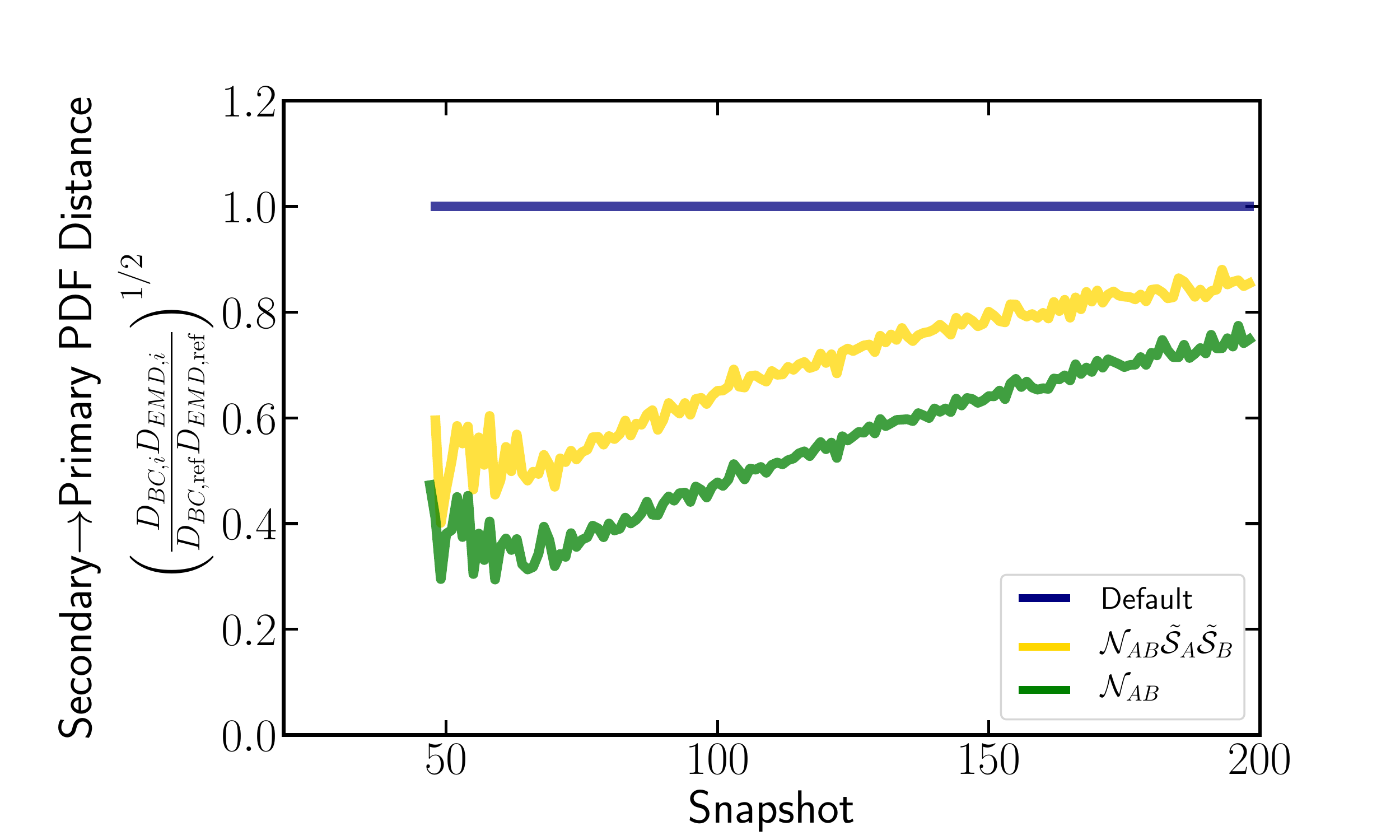}
    \caption{{\bf Merit Function Comparison:} We show the ratio of PDF distances between our default merit and two other merit functions. Smaller values indicate worse separation between primary and secondary links.}
    \label{fig:meritcomp}
\end{figure}
The ideal merit function is one which well separates primaries and secondaries, or more formally, the merit distribution of primary links differs significantly from that of secondary links. We can compare distributions in a variety of ways. One commonly used measure of the similarity of probability distribution functions is the Bhattacharyya distance, $D_{BC}$, which is related to the Bhattacharyya coefficient, $BC$ \cite[][]{bhattacharyya1943}. For probability distributions $p$ and $q$ defined over a domain $X$, the coefficient and distances are: 
\begin{align}
    BC (p, q) &= \sum_{x\in X}{\sqrt {p(x)q(x)}},\\
    D_{BC}(p, q) &= -\ln \left(BC(p,q)\right).
\end{align}
Distributions with no overlap have $D_{BC}=\infty$. 

\par
Another common distance measure is the $1^{\rm st}$ Wasserstein distance or so-called Earth Mover's Distance, which measures the minimum work needed to be done to transform $p\rightarrow q$ \cite[e.g.][]{rubner1998a}. Consider a set $P= (p_1, p_2, \dots, p_m)$ and $Q = (q_1, q_2, \dots, q_n)$ with a set of distances $D = [d_{i,j}]$, where $d_{i,j}$ is the ground distance between $p_i$ and $q_j$. The earth mover's distances $D_{EMD}$ is determined by finding the optimal flow $F=[f_{i,j}]$ that minimises the overall cost of moving set $P$ to $Q$, 
\begin{align}
    \min {\sum _{i=1}^{m}\sum _{j=1}^{n}f_{i,j}d_{i,j}},
\end{align}
subjected to the constraints:
\begin{gather}
    f_{i,j}\geq 0,1\leq i\leq m,1\leq j\leq n, \\
    \sum _{j=1}^{n}{f_{i,j}}\leq w_{pi},1\leq i\leq m, \\
    \sum _{i=1}^{m}{f_{i,j}}\leq w_{qj},1\leq j\leq n, \\
    \sum _{i=1}^{m}\sum _{j=1}^{n}f_{i,j}=\min \left\{\ \sum _{i=1}^{m}w_{pi},\quad \sum _{j=1}^{n}w_{qj}\ \right\}.
\end{gather}
The earth mover's distance is defined as the work normalised by the total flow:
\begin{align}
    D_{EMD}=\frac {\sum _{i=1}^{m}\sum _{j=1}^{n}f_{i,j}d_{i,j}}{\sum _{i=1}^{m}\sum _{j=1}^{n}f_{i,j}}.
\end{align}

\par 
The merit function should produce the maximum distance between primary and secondary distributions. We show the product of the Bhattacharyya distance and the Wasserstein distance in \Figref{fig:meritcomp}. This figure clearly shows how using ranking can improve the separation between primaries and secondaries and how using only the most bound particles can significantly improve classification.

\section{Configuration Options}
\label{sec:appendix:tables}
We list the complete configuration options here.
{
\onecolumn
%\begin{table*}
%\setlength\tabcolsep{2pt}
\centering\footnotesize
\begin{longtable}{@{\extracolsep{\fill}}p{0.125\textwidth}|p{0.2\textwidth}|c|p{0.5\textwidth}}
\caption{\textsc{TreeFrog} configuration parameters}
\label{tab:treefrog:config}\\

%\begin{table*}
%\setlength\tabcolsep{2pt}
%\centering\footnotesize
%\label{tab:treefrog:config}
%\begin{tabular}{@{\extracolsep{\fill}}p{0.125\textwidth}|p{0.2\textwidth}|c|p{0.5\textwidth}}

\hline
\hline
    & Name & Default Value & Comments\\
    \hline
    Base Tree Construction Options & & & Related to basic operation. \\
    \hline
    & Tree\_direction & 1
        & Integer indicating direction in which to process snapshots and build the tree. Descendant [1], Progenitor [0], or Both [-1].\\
    & Particle\_type\_to\_use & -1
        & Particle types to use when calculating merits. All [-1], Gas [0], Dark Matter [1], Star [4].\\
    & Default\_values& -1
        & Whether to use default cross matching \& merit options when building the tree. 1/0 for True/False.\\

    \hline
    Input/Output Options & & & Related to input/output formats. \\
    \hline
    & Input\_tree\_format & 2
        & Integer flag indicating input halo catalogue format. ASCII SUSSING format \cite[see][]{srisawat2013} [1], VELOCIraptor catalogues [2], ASCII nIFTy format, ASCII VOID catalogue format \cite[see][]{sutter2014a}. \\
    & VELOCIraptor\_input\_& & \\   
    & format & 2
        & Integer flag indicating input format of \textsc{VELOCIraptor} catalogue. ASCII [0], Binary [1], HDF5 [2]. \\
    & VELOCIraptor\_input\_& & \\
    & field\_sep\_files & 0
        & Flag indicating whether halos and subhalos are written in separate files. All (sub)halos together [0], separate [1].\\
    & VELOCIraptor\_input\_& & \\
    & num\_files\_per\_snap & 1
        & If VELOCIraptor run in MPI mode, more than one file produced. Multiple files [1], one files [0].\\
    & Output\_format & 2
        & Integer flag for output format. ASCII [0], HDF5 [2].\\
    & Output\_data\_content & 1
        & Integer flag for data contained in the output. BASIC (only descendant or progenitor IDs) [0], Standard (IDs plus merit) [1], Verbose (IDs, merit, and number of particles in structure). \\

    \hline
    Merit Options & & & Related to calculation of merit function. \\
    \hline
    & Merit\_type & 6  
        & Integer specifying merit function to use. Optimal descendant tree merit in \Eqref{eqn:merit} [6], common (progenitor tree) merit in \Eqref{eqn:nshared} [1].\\
    & Core\_match\_type & 2
        & Integer flag indicating the type of core matching used. Off [0], core-to-all [1], core-to-all followed by core-to-core [2], core-to-core only [3].\\
    & Particle\_core\_fraction & 0.4
        & Fraction of particles to use when calculating merits. Assumes some meaningful rank ordering to input particle lists and uses the first $f_{\rm TF}$ fraction.\\
    & Particle\_core\_min\_ & &\\
    & numpart & 5
        & Minimum number of particles to use when calculating merit if core fraction matching enabled.\\
    & Shared\_particle\_signal\_to\_ & & \\
    & noise\_limit & 1
        & Mininum significance $\sigma_N$ of number of shared particles between object $i$ and $j$, such that links with $N_{i\bigcap j}<\sigma_N \sqrt{N_i}, N_{i\bigcap j}<\sigma_N \sqrt{N_j}$, that is where number of shared particle is below Poisson fluctuations, are removed. \\

    \hline
    Temporal Linking Options & & & Related to how code searches for candidate links across multiple snapshots.\\
    \hline
    & Nsteps\_search\_new\_links & 1
        & Number of snapshots to search for links.\\
    & Multistep\_linking\_criterion & 3
        & Integer specifying the criteria used when deciding whether more snapshots should be searched for candidate links. Criteria depend on tree direction. {\bf Descendant Tree:} continue searching if halo is: missing descendant [0]; missing descendant or descendant merit is low [1]; missing descendant or missing primary descendant [2]; missing a descendant, a primary descendant or primary descendant has poor merit [3]. {\bf Progenitor tree}: [0,1]. \\
    & Merit\_limit\_continuing\_ & & \\
    & search & 0.025 
        & Float specifying the merit limit a match must meed if using Multistep\_linking\_criterion=[1,3].\\
    & Temporal\_merit\_type & 1
        & Integer specifying how merits at different times are weighted. {\bf Descendant Tree}: Adjusts weights according to ranking and ignore temporal information for descendant trees [1], Adjust weights using ranking and temporal information [0]. {\bf Progenitor Tree}: Adjust weights using temporal information [0]. \\
    & Merit\_ratio\_limit & 4.0
        & For objects with secondary descendants but no primary descendant where secondary descendant's primary progenitor also possibly primary progenitor of another object, maximum merit ratio between the secondary descendant's primary progenitor and current object for which ranking is altered, leaving secondary now primary and previous primary now primary descendant of different object. \\
    
    \hline
    Additional Options & & & \\
    \hline
    & Max\_ID\_Value & 134217728
        & \textsc{TreeFrog} assumes particle IDs range from [0,MaxID] and uses this information for internal indexing. Set this value or invoke some form of mapping that maps input IDs to this form. Code will allocate memory of size MaxID to quickly access particle group ids. \\
    & Mapping & 0 
        & Integer specifying the type of mapping to use on input particle IDs. No mapping [0], generate a id to index map (computationally intensive) [-1], or [1] a user defined mapping. If number of particles is large, suggestion is to invoke [-1]. This needs to only be done once and the code will save the map. \\
    & Temporal\_haloidval & 1000000000000
        & For temporally unique halo IDs, \\
    & HaloID\_snapshot\_offset & 0
        & Offset applied to all temporal halo id values. Halo IDS have added to them (input snapshot number+HaloID\_snapshot\_offset)*Temporal\_haloidval. \\
    & HaloID\_offset & 0
        & Offset applied to all halo id values. Halo IDS are then input index+1+HaloID\_offset. \\
\end{longtable}
%\end{tabular}
%\end{table*}
}
\twocolumn

\section{Associated Tools}
\label{sec:appendix:examples}
\textsc{TreeFrog} comes with a {\sc Python-2/3} tool-kit, specifically routines to manipulate the output data produced by the various codes. Typically, these produce {\sc dict} containing {\sc numpy} arrays, allowing for quick analysis and plotting. The repositories also come with examples of producing metric plots. The codes are  {\sc Python-3} (compatible with {\sc Python-2}) and make use of {\sc numpy}, \textsc{h5py}, \textsc{scipy}, \textsc{matplotlib}, and \textsc{scikit.learn}.

\end{appendix}

\end{document}